\documentclass[sigplan,screen]{acmart}

\copyrightyear{2024}
\acmYear{2024}
\setcopyright{rightsretained}
\acmConference[ASPLOS '24]{29th ACM International Conference on Architectural Support for Programming Languages and Operating Systems, Volume 3}{April 27-May 1, 2024}{La Jolla, CA, USA}
\acmBooktitle{29th ACM International Conference on Architectural Support for Programming Languages and Operating Systems, Volume 3 (ASPLOS '24), April 27-May 1, 2024, La Jolla, CA, USA}
\acmDOI{10.1145/3620666.3651380}
\acmISBN{979-8-4007-0386-7/24/04}

\usepackage{packages}
\usepackage{macros}

\usepackage[]{hyperref}

\begin{document}

\title{\neupims: NPU-PIM Heterogeneous Acceleration for Batched LLM Inferencing}

\author{
  \hfill
  Guseul Heo\hfill
  Sangyeop Lee\hfill
  Jaehong Cho\hfill
  Hyunmin Choi\hfill
  Sanghyeon Lee\hspace*{\fill} \\
  \hfill
  Hyungkyu Ham\footnotemark[2]\hfill
  Gwangsun Kim\footnotemark[2]\hfill
  Divya Mahajan\footnotemark[4]\hfill
  Jongse Park\hspace*{\fill}
}

\affiliation{%
  \vspace{1ex}
  \institution{KAIST, \footnotemark[2]POSTECH, \footnotemark[4]Georgia Institute of Technology}
  \country{}
  \vspace{1ex}
}

\affiliation{%
    \customopen\customemail{gsheo,} 
    \customemail{sangyeop-lee,}\customspace
    \customemail{jhcho,}\customspace
    \customemail{hmchoi,}\customspace
    \customemail{leesh6796}\customclose
    \customemail{@casys.kaist.ac.kr}
    \country{}
}
\affiliation{%
    \hfill
    \customopen \customemail{hhk971,}\customspace
    \customemail{g.kim}\customclose
    \customemail{@postech.ac.kr} \quad \customemail{divya.mahajan@gatech.edu} \quad
    \customemail{jspark@casys.kaist.ac.kr}\hspace*{\fill}
    \country{}
}

\renewcommand{\authors}{Guseul Heo, Sangyeop Lee, Jaehong Cho, Hyunmin Choi, Sanghyeon Lee, Hyungkyu Ham, Gwangsun Kim, Divya Mahajan and Jongse Park}
\renewcommand{\shortauthors}{Heo et al.}

\vspace{2ex}
\begin{abstract}
Modern transformer-based Large Language Models (LLMs) are constructed with a series of decoder blocks.
Each block comprises three key components: (1) QKV generation, (2) multi-head attention, and (3) feed-forward networks.
In batched processing, QKV generation and feed-forward networks involve compute-intensive matrix-matrix multiplications (GEMM), while multi-head attention requires bandwidth-heavy matrix-vector multiplications (GEMV).
Machine learning accelerators like TPUs or NPUs are proficient in handling GEMM but are less efficient for GEMV computations.
Conversely, Processing-in-Memory (PIM) technology is tailored for efficient GEMV computation, while it lacks the computational power to handle GEMM effectively.

Inspired by this insight, we propose \neupims, a heterogeneous acceleration system that jointly exploits a conventional GEMM-focused NPU and GEMV-optimized PIM devices.
The main challenge in efficiently integrating NPU and PIM lies in enabling concurrent operations on both platforms, each addressing a specific kernel type.
First, existing PIMs typically operate in a ``blocked'' mode, allowing only either NPU or PIM to be active at any given time.
Second, the inherent dependencies between GEMM and GEMV in LLMs restrict their parallel processing.
To tackle these challenges, \neupims is equipped with \emph{dual row buffers} in each bank, facilitating the simultaneous management of memory read/write operations and PIM commands.
Further, \neupims employs a runtime \emph{sub-batch interleaving} technique to maximize concurrent execution, leveraging batch parallelism to allow two independent sub-batches to be pipelined within a single \neupims device.
Our evaluation demonstrates that compared to GPU-only, NPU-only, and a na\"ive NPU+PIM integrated acceleration approaches, \neupims achieves 3$\times$, 2.4$\times$ and 1.6$\times$ throughput improvement, respectively.

\end{abstract}



\maketitle 

\section{Introduction}
\label{sec:intro}

\begin{figure*}[t]
\centering
\includegraphics[width=0.95\textwidth]{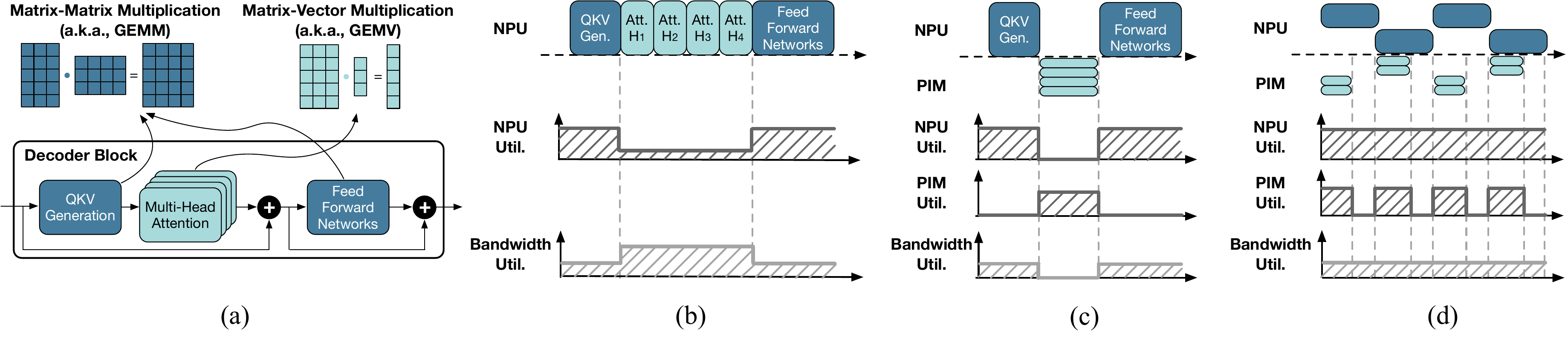}
\vspace{-3ex}
\caption{(a) Mathematical components of decoder blocks that constitute LLMs, (b) NPU-only baseline accelerator equipped with \emph{non}-PIM memory (e.g., GPU), (c) NPU+PIM integrated baseline accelerator, and (d) the proposed \neupims accelerator.}
\vspace{-2ex}
\label{fig:intro-overview}
\end{figure*}

Large Language Models (LLMs) are being widely deployed across various sectors such as natural language understanding~\cite{openai23-gpt4,google23-palm2,zeng23-glm130b,hoffmann22-chinchilla,black22-gptneox20b,zhang22-opt,touvron23-llama}, content generation~\cite{dalle2,chen21-llm-code,meta23-code-llama,copilot}, and decision support~\cite{decision_llm}.
However, a key challenge with these models is the substantial resource requirement they impose - both memory and compute.
This paper specifically addresses the inference challenges in contemporary LLMs, with an emphasis on models like GPT4~\cite{openai23-gpt4} and LLaMA~\cite{touvron23-llama}.

The algorithmic commonality of these state-of-the-art LLMs is that their model architecture constitutes a stack of decoder blocks.
As illustrated in Figure~\ref{fig:intro-overview}(a), each block is structured around three primary layers: (1) Query-Key-Value (QKV) generation, (2) Multi-Head Attention (MHA), and (3) Feed-Forward Networks (FFNs).
For efficient computation of these blocks, a prevalent strategy is batching multiple inference requests. 
Batching allows QKV generation and feed-forward layers to reuse weights across multiple requests, resulting in GEneral Matrix Multiplication (GEMM) operations between weight and activation matrices.
Conversely, the multi-head attention layer requires multiplication between activation matrices and activation vectors with no data reuse opportunity, leading to GEneral Matrix-Vector Multiplication (GEMV) operations.

Overall, LLM inference involves the computation of numerous large-scale GEMMs and GEMVs.
To address this computational demand, a common practice is to utilize high-performance machine learning (ML) accelerators, such as GPUs and TPUs. In this paper, we will refer to these ML accelerators as Neural Processing Units (NPUs).
NPUs are often optimized for compute-intensive tasks, particularly for the efficient execution of GEMMs. 
However, their utility for GEMVs is less optimal due to the latter's lower arithmetic intensity, which leads to under-utilization of the NPU's computational resources.
On the other hand, Processing-in-Memory (PIM) technology~\cite{9869326, 9912078, 9912072, 10.1145/3547353.3522661, 10.1145/3489048.3522661, 9771457,AttAcc, 9365862, hcs23-aquabolt-xl, 10.1145/3490422.3502355, hbm_pim,newton,tensor-dimm, RecNMP, SpaceA, gradpim, aespa,10.1145/3458817.3476146, 9474146, infstream, 9563039,computedram, dual-micro, floatpim, pimcloud, 7284059, recross,lpddr-gpt,pim-instr, graphpim, virtualpim, chameleon-pim, 9566881, heo2023primo, hyun2023pathfinding, jonatan2024scalability, micro21-trim}, while not as effective for GEMMs, shows promise for the bandwidth-intensive GEMV operations.

To this end, this work proposes \neupims, a novel heterogeneous acceleration system for batched inference of LLMs. 
We architect \neupims such that it effectively balances the utilization of memory bandwidth and computational resources of the system to improve the overall inference throughput.
\neupims jointly exploits (1) a conventional GEMM-centric NPU using a 2D cluster of multiple systolic arrays and (2) a multitude of GEMV-friendly processing-in-memory (PIM) accelerators. 
In designing \neupims, we identify two major challenges: 
\begin{itemize}[labelindent=0.1em,nolistsep,leftmargin=1em]
\item \textbf{Microarchitectural Challenge:} Current PIMs operate in a ``blocked'' mode, preventing the simultaneous execution of NPU and PIM. This serialization leads to an inherent under-utilization of resources.
\item \textbf{Algorithmic Challenge:} In LLM decoder block, GEMM and GEMV operations have a data dependency. This algorithmic limitation fundamentally limits the possibility of parallel NPU+PIM computations.
\end{itemize}

\neupims addresses the aforementioned challenges by taking a hardware-algorithm co-design approach and makes the following contributions:

\niparagraph{(1) Microarchitectural Contribution: }
To facilitate NPU+PIM parallel execution, \neupims introduces a modified PIM bank architecture that enables regular memory accesses to occur concurrently with GEMV operations within the PIM. 
This is achieved by employing distinct row buffers for these two functionalities, hereafter referred to as \textit{dual row buffers}.
Dual row buffers leverage the property of DRAM where multiple rows can be activated independently without affecting functionality.
This further requires handling and scheduling of mixed commands for memory access and PIM operation at the memory controllers without violating DRAM timing parameters.
To do so, \neupims strategically intersperses the two types of commands, minimizing row activation delays. Additionally, a few composite commands are appended to the baseline PIM ISA, performing multiple GEMV operations and thereby amortizing the controlling cost.

\niparagraph{(2) Algorithmic Contribution: }
To enable parallel executions of GEMM and GEMV operators within the decoder block, we introduce the \emph{sub-batch interleaving} technique, which concurrently processes two sub-batch inference computations on the \neupims system.
As the two sub-batches are independent of each other, it is possible to parallelize the execution of GEMM operations from one sub-batch with GEMV operations from another sub-batch.
This approach creates avenues for simultaneous executions, enhancing overall efficiency.
With sub-batch interleaving, \neupims aims to balance the workload between GEMM and GEMV operations effectively.
To balance the pipeline of sub-batches, we estimate mappings from sequence lengths to the MHA execution latency on the PIM. 
This information allows \neupims to partition a given batch such that it balances the total sum of sequence lengths in the sub-batches.

Combining the proposed microarchitectural and algorithmic innovations, \neupims achieves high utilization on both NPU and PIM accelerators, thus offering significant throughput improvement over NPU-only and na\"ive NPU+PIM integrated baselines. 
Figure~\ref{fig:intro-overview}(b)-(d) visualizes the operator-accelerator mappings on NPU and/or PIM, along with their utilization trends for a short window in the execution runtime.

We evaluate the effectiveness of \neupims using 4 variants of GPT3, a state-of-the-art LLM, with varying sizes. The evaluation utilizes real-world LLM inference datasets, ShareGPT and Alpaca, both accompanied by input and output sequence length information.
We develop the \neupims simulator\footnote{Our simulator is available at \url{https://github.com/casys-kaist/NeuPIMs}.} by integrating an open-source NPU simulator, ONNXim~\cite{onnxim}, with our in-house PIM simulator built on DRAMsim3~\cite{dramsim3}.
Our experimental results report that compared to an NPU-only and a na\"ive NPU+PIM integrated baseline accelerators, \neupims achieves 2.4$\times$ and 1.6$\times$ throughput improvement, respectively.
These significant throughput gains are attributed to the improved resource utilization of NPU and PIM from 28\% and 17\% to 65\% and 26\%, respectively.
These compelling advantages highlight that \neupims effectively overcome the limitations of existing solutions and take an effective initial step towards the practical deployment of PIM for LLM inference scenarios.
\section{Background}
\label{sec:background}

\subsection{Computational Characteristics of LLM Inference}
\label{sec:llm-background}

\begin{figure}[t]
\centering
\includegraphics[width=1.0\linewidth]{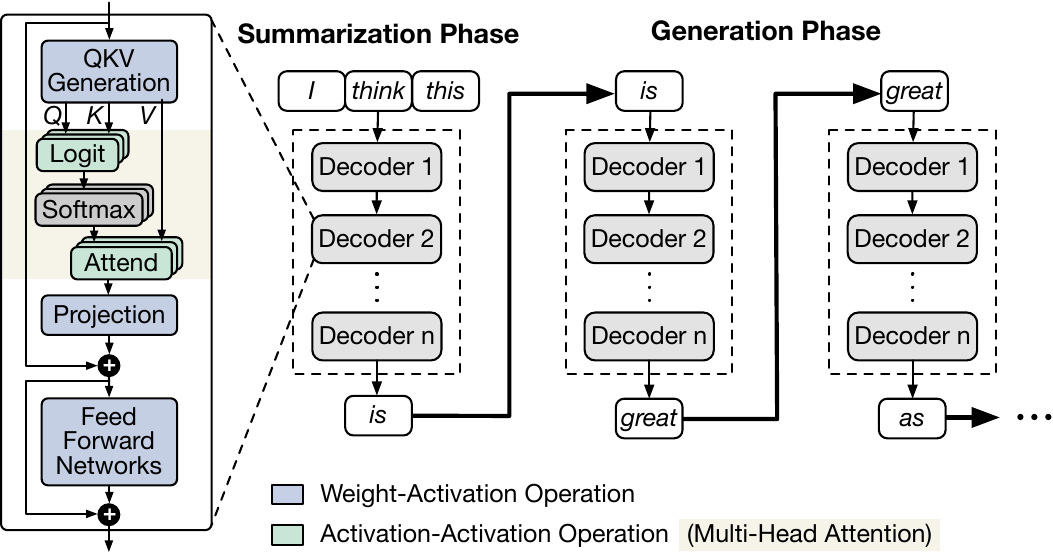}
\vspace{-4ex}
\caption{Model architecture and inference in LLMs.}
\vspace{-2ex}
\label{fig:llm-inference}
\end{figure}

\niparagraph{Model architecture and execution of LLMs.}
Figure~\ref{fig:llm-inference} illustrates the model architecture that all state-of-the-art large language models share~\cite{bert:2018, touvron23-llama, zhang22-opt, mpt, google23-palm2, megatronlm}. 
This illustration serves as a recurring example throughout the paper.
For an input prompt (e.g., "I think this"), the model undergoes an summarization phase, encoding the input to establish context for the subsequent generation phase.
In the generation phase, the model produces tokens one at each iteration in an autoregressive manner, using the generated key-value projections for the next iteration.
Both phases constitute a sequence of decoder blocks, each comprising three major layers: (1) QKV generation, (2) multi-head attention (MHA), and (3) a set of feed-forward networks (FFNs).

\begin{figure}[t]
\centering
\includegraphics[width=0.9\linewidth]{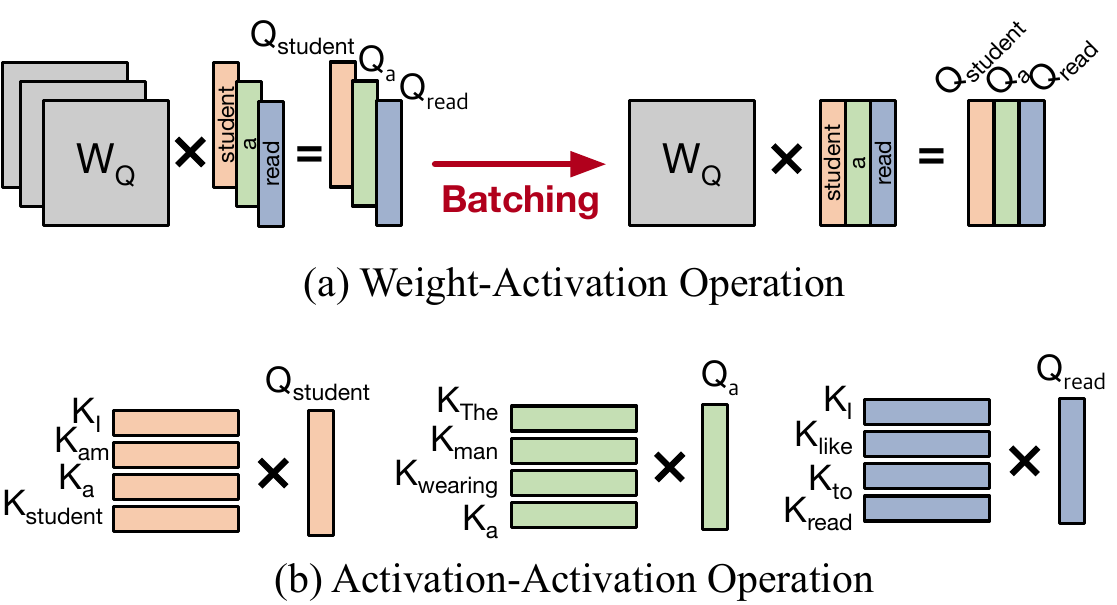}
\vspace{-3ex}
\caption{Operators in a LLM decoder block.}
\vspace{-2ex}
\label{fig:llm-batching}
\end{figure}

\niparagraph{Batched inference of LLMs.}
Computationally, the MHA layers have significantly different characteristics than QKV generation and FFN layers. 
Figure~\ref{fig:llm-batching} denotes the example tensor operations of LLMs that show (a) weight-activation multiplications, and (b) activation-activation multiplications. 
The computations for QKV generation and FFNs are performed by multiplying a per-token Q/K/V activation or attention vector with the trained weight matrices (GEMV). 
However, these GEMV operators are transformed into GEMMs when (1) they are located at the decoders in the summarization phase, getting multiple token vectors in parallel (e.g., ``I think this'') or (2) multiple inferences are batched, further parallelizing the computations for multiple single-token generation processes. 
On the other hand, the computations for MHA layers are multiplications between two different activations where one is for the current token (vector), and the other is for all the tokens before the current token (matrix), rendering a matrix-vector multiplication (GEMV). 
As the activation operands are unique for each inference request, their batching is not possible, making the computations highly memory bandwidth-bound.  

\begin{figure}[t]
\centering
\includegraphics[width=1.0\linewidth]{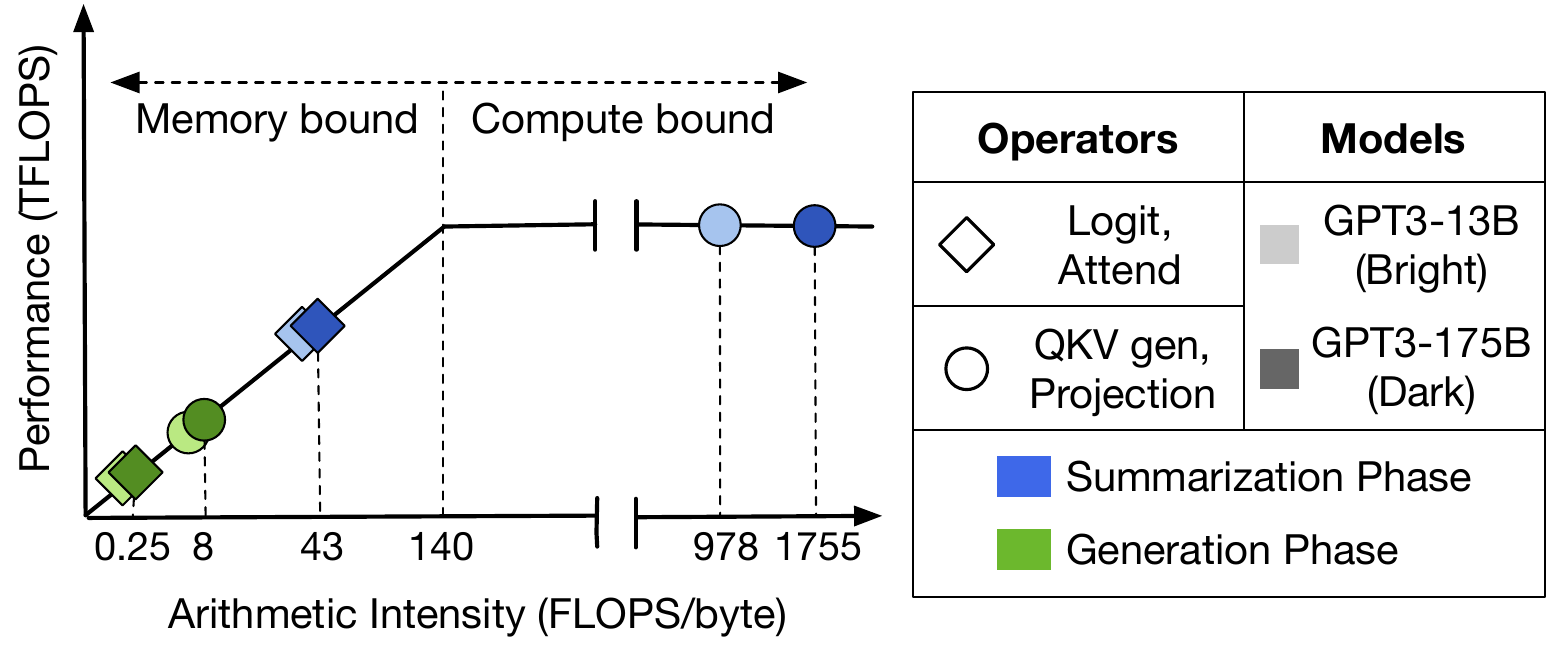}
\vspace{-4ex}
\caption{Arithmetic intensities of LLM layers.}
\label{fig:arithmetic-intensity}
\vspace{-3ex}
\end{figure}

\niparagraph{Analysis of arithmetic intensity.}
To better understand the computational characteristics of LLM inference, we conduct a roofline analysis using two GPT3 variants, GPT3-13B and GPT3-175B. 
Figure~\ref{fig:arithmetic-intensity} shows the relationship between the arithmetic intensity (FLOPS/byte) and performance (TFLOPS).
We observe that for both models, the generation phases are severely memory-bound, while the summarization phases are compute-bound.
As these two phases have algorithmic dependencies and occur alternately in a sequential manner, it is fundamentally challenging to achieve high resource utilization using a homogeneous computing platform.   
This insight motivates this work and drives us to design a heterogeneous system that combines a compute-centric systolic array-based NPU for GEMMs with memory-centric Processing-in-Memory (PIM) accelerators for GEMVs.

\subsection{LLM Inference Serving}
As there is a massive resource demand for LLMs, the de-facto practice is to build large-scale inference serving frameworks such as DeepSpeed~\cite{deepspeed}, Orca~\cite{orca} and vLLM~\cite{vllm}.
These frameworks offer inference services for customer requests (i.e., prompts), which enables batching.

\niparagraph{Selective batching.}
In general, batching is an effective method for neural network inference to improve resource utilization, while not sacrificing the latency requirement.
However, MHA layers pose a challenge as they do not allow batching. 
This presents a difficulty for hyperscalers dealing with numerous customer requests while operating within limited compute resources.
To address this, a recent work, Orca~\cite{orca}, proposes a solution where attention layers are individually computed, while QKV generation and FFN layers are batched. 
This approach allows the system to still benefit from batching when possible; otherwise, it serializes the computation.
This unique algorithmic property necessitates the simultaneous computations of GEMM and GEMV, which is the main motivation for this work.

\niparagraph{Iteration level scheduling.}
Inference serving system receives requests in a streaming fashion without a deterministic schedule. 
Therefore, there is a need for a parallelization approach that can efficiently process the non-deterministically collected set of inference requests.
Orca~\cite{orca} additionally proposes to schedule batched inference at the beginning of every iteration. 
This allows new inference requests to be added to and terminated requests to be removed from the batch. 
Consequently, newly arrived requests do not need to wait until the generation phase for an already-started batch is terminated.
This approach can significantly reduce the average latency for inference serving.
\neupims is built upon this scheduling technique, and thus, it manages the inference requests at the iteration boundaries. 

\niparagraph{Memory paging for attention.} 
vLLM~\cite{vllm} is another recent effort to enhance the resource utilization of LLM inference serving systems, with a specific focus on memory management.
As discussed in Section~\ref{sec:llm-background}, the QKV generation layer produces KV cache, the input for the attention layers that can be reused in the generation phases. 
Leveraging this opportunity, LLM inference systems \emph{cache} the KV projections in the memory, the size of which can be significant when the sequence length becomes large. 
vLLM introduces memory paging for this cached data, ensuring that a significant amount of memory is not pre-allocated long before its actual use.
\neupims employs the vLLM's paging technique, implementing the page-based memory allocation mechanism for KV cache, which effectively increases the batch size significantly. 

\emph{It is worthwhile to note that \neupims is designed to be deployed on an inference serving system that incorporates all of these aforementioned techniques.}

\section{Motivation}
\label{sec:motiv}
This section provides the motivation that underlies the design decisions of \neupims.
First, we identify problems of the existing GPU-based LLM inference serving systems, which motivates the NPU-PIM heterogeneous approach. 
Then, we will discuss the limitations of na\"ive NPU-PIM integration approach, defining the target research challenges of this work.

\subsection{GPU-based LLM Inference Serving}
\label{sec:gpu-analysis}

Particularly for LLMs, as they require a lot of memory, it is a common practice to deploy them on a cluster of multiple GPUs~\cite{deepspeed, nvidia_triton, tensorrt, onnxruntime, hugging-face}, leveraging pipeline parallelism~\cite{pipedream, gpipe} and/or tensor parallelism~\cite{megatronlm, alpa}.

\begin{figure}[t]
        \centering
        \includegraphics[width=0.9\linewidth]{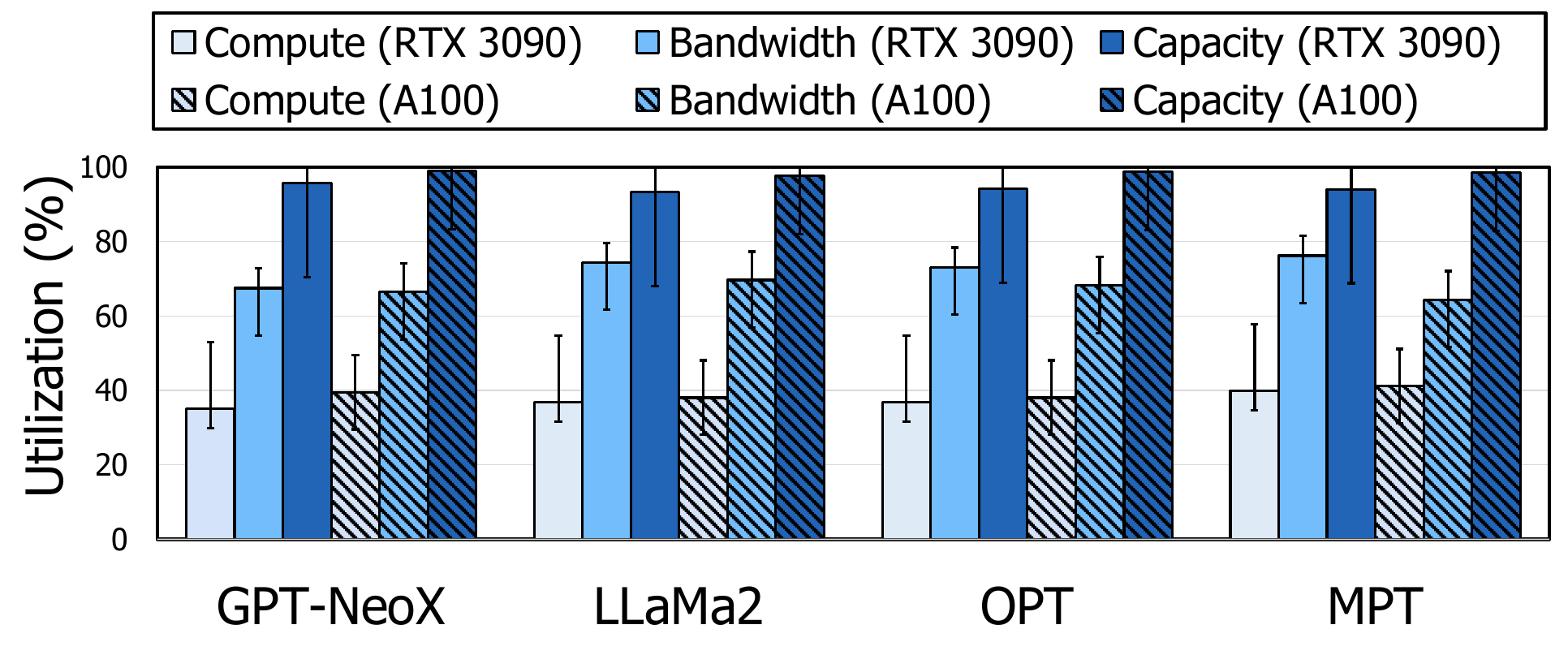}
        \vspace{-2ex}
        \caption{GPU resource utilization for four different LLMs. }
        \vspace{-3ex}
        \label{fig:gpu-utilization}
\end{figure}

\niparagraph{Under-utilization of the GPU system.}
We analyze a GPU-equipped baseline system to understand the utilization of compute, memory, and bandwidth for LLM inference.
We compare systems with NVIDIA GeForce RTX 3090 24GB and NVIDIA A100 40GB, running four different LLM models: GPT-NeoX, LLaMa2, OPT, and MPT. 
Figure~\ref{fig:gpu-utilization} presents the utilization results along with the layer-wise variations as error bars. 
The figure illustrates that the capacity utilization closely approaches 100\% despite the inherent imperfections in the parallelization schemes.
This observation is intuitive as the number of GPUs used is determined based on the capacity constraints.
However, the utilization of computational resources is consistently lower than 40\%, which shows the cost-ineffectiveness of GPU-based LLM inference systems. 
This under-utilization is attributed to insufficient bandwidth, despite A100s being equipped with HBM, providing an aggregate of 1,555 GB/s.
Unfortunately, this imbalance is inevitable as long as serial dependencies between GEMMs and GEMVs persist. 

\begin{figure}[t]
        \centering
        \includegraphics[width=0.9\linewidth]{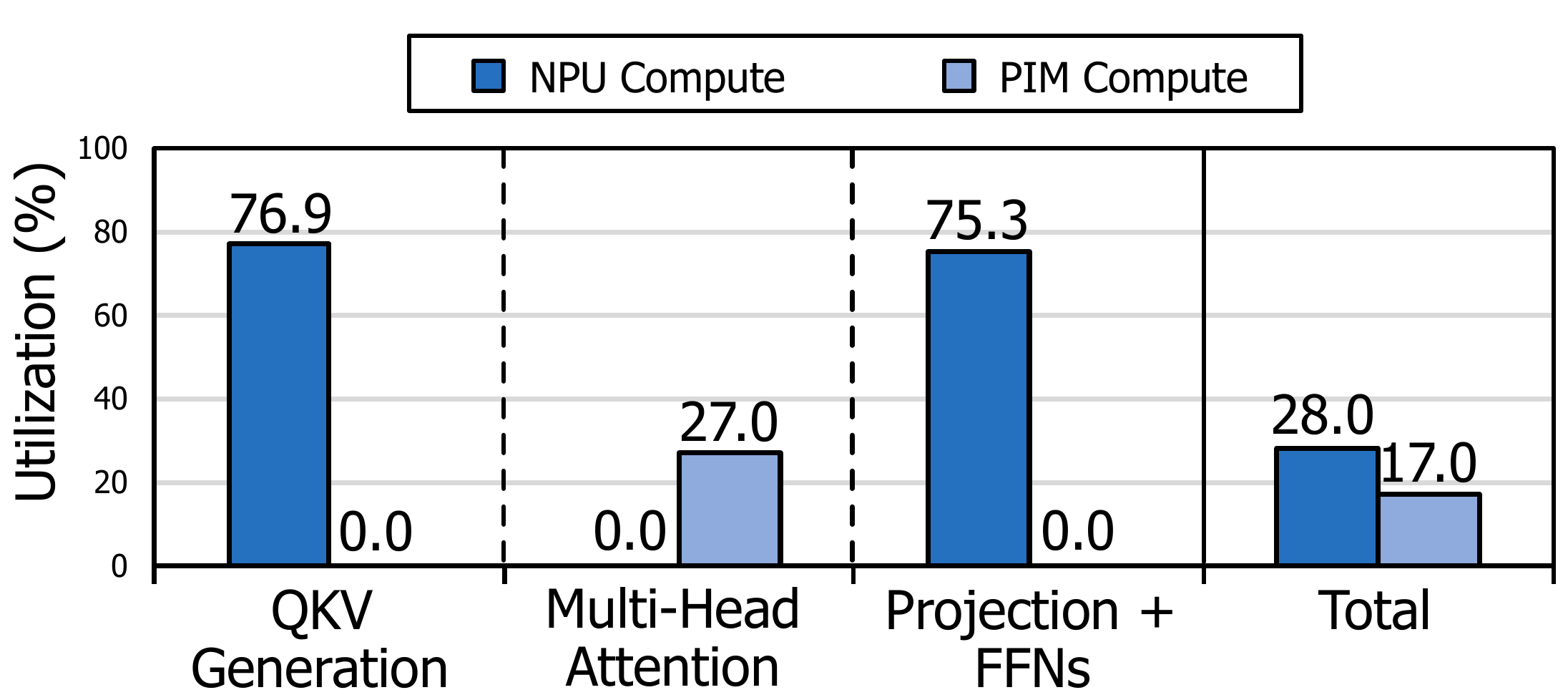}
        \vspace{-2ex}
        \caption{NPU-PIM resource utilization for decoder block.}
        \vspace{-3ex}
        \label{fig:npupim-utilization}
\end{figure}

\begin{figure*}[t]
        \centering
        \includegraphics[width=1.0\textwidth]{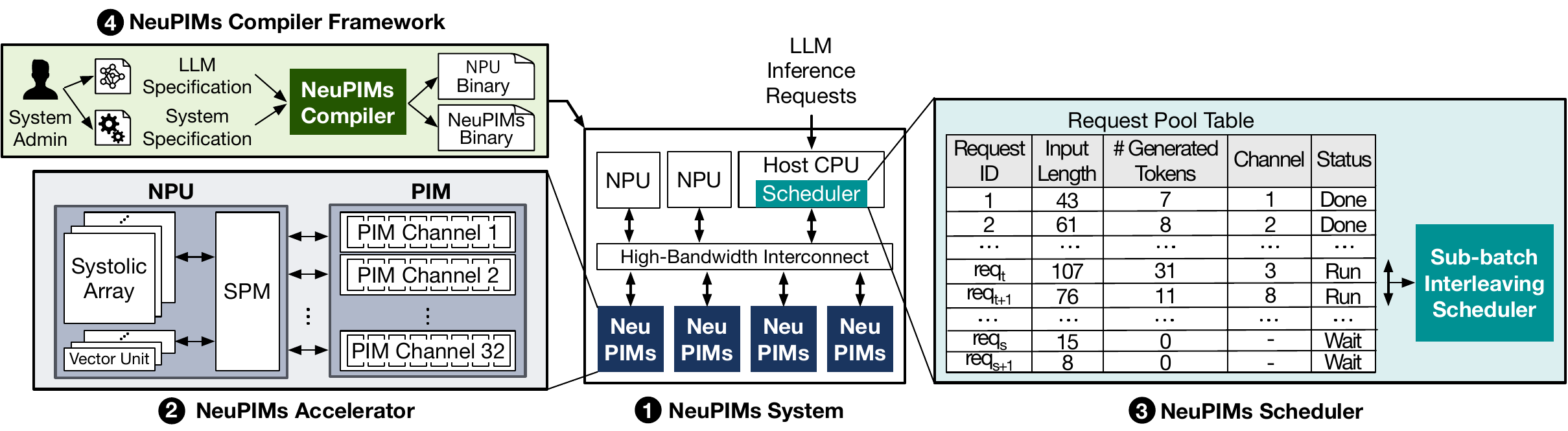}
        \vspace{-3ex}
        \caption{Overview of the proposed \neupims system.}
        \vspace{-2.5ex}
        \label{fig:overview}
\end{figure*}

\subsection{A Na\"ive NPU-PIM Approach}

A straightforward approach to resolve this bandwidth bottleneck is to exploit the Processing-in-Memory (PIM) technology, which allows offloading the bandwidth-bound computations to its in-memory accelerator.
Thus, we design a na\"ively integrated NPU-PIM accelerator, exploiting a systolic array architecture for the NPU and incorporating a state-of-the-art PIM-based GEMV accelerator, Newton~\cite{newton}.
We use the same methodology as in Section~\ref{sec:gpu-analysis}, where we replace GPUs with a standard NPU-PIM integrated device. 
The detailed hardware and system simulation methodology are described in Section~\ref{sec:method}.
Figure~\ref{fig:npupim-utilization} presents the compute utilization of NPU and PIM for running different layers in the LLM decoder blocks.
The results show that while NPU is busy running QKV generation, projection, and FFN layers, PIM utilization stays at zero. 
On the other hand, NPU utilization becomes almost zero when PIM is running the MHA layers.
Consequently, the combined utilization of NPU and PIM, when measured across the entire execution time, is less than 40\% for both.

\niparagraph{Necessity of concurrent NPU and PIM executions.}
The observed under-utilization is primarily due to the fundamental limitation in PIM's microarchitecture that disallows the concurrent execution of host (NPU) and PIM units, which serializes the disjoint resource usages.  
Consequently, the most critical challenge in realizing the practical use of PIM in NPU accelerators is enabling their parallel executions. 
This research problem constitutes the focus of this work.
\section{Overview of \neupims}
\label{sec:neupims}

Figure~\ref{fig:overview} illustrates the overview of the proposed \neupims system.
This system alleviates the low resource utilization of an LLM inference serving system.
To achieve this goal, \neupims comprises: (1) an NPU equipped with systolic arrays, vector processing units, and multiple HBM-based PIM channels that collaboratively process the batched inference requests, and (2) a scheduler that partitions an inference batch into two sub-batches and leverages sub-batch parallelism to enable their interleaved executions for enhanced NPU-PIM parallelization. 
Note that NPU in the \neupims device is a general representation of any ML accelerator such as TPU~\cite{tpuv4} and is not the contribution of this work.
In this paper, we propose a novel accelerator integrating PIM with NPU and the corresponding scheduling strategy.
First, we provide an overview of the system. 

\niparagraph{\circled{1} \neupims system.}
This system comprises a host CPU, multiple \neupims devices (i.e., NPU+PIM accelerators), and multiple standalone NPUs connected through a high-bandwidth interconnect such as PCIe and CXL.
As the summarization phase is entirely composed of GEMMs, we delegate its computation to the standalone NPUs, while \neupims focuses on the computation of the generation phase. 
While the diagram visualizes a single-node \neupims system, the system can scale to multiple nodes, which will be discussed in more detail in Section~\ref{sec:scale}.
As in typical inference serving systems, our system receives the LLM inference requests in a streaming fashion. 
The requests are assigned to a PIM channel and queued in the request pool table until the on-going iteration is completed and a new iteration commences for the execution. 

\niparagraph{\circled{2} \neupims accelerator.}
We extend the design of standard PIM to support \neupims LLM inferencing strategy. 
The bank architecture is expanded to include dual row buffers—one for PIM execution and the other for regular memory accesses, as illustrated in Figure~\ref{fig:newton-vs-neupims}.
The dual row buffer architecture enables the NPU to perform memory read/write accesses on the bank rows that are not currently in use for PIM computations. This segregation is regulated by memory controllers to ensure that multiple activations are not issued over the same bank row (Section~\ref{sec:architecture}).

\niparagraph{\circled{3} \neupims scheduling algorithm.} 
Our prototype \neupims accelerator has 32 HBM-based PIM channels, each of which is controlled by its own memory controller. 
The memory controllers manage the interleaving of memory read/write commands and PIM commands in a way that the inter-command timing delays are not violated, while maximizing the control path throughput. 
Effective interleaving is critical for performance since it directly affects the concurrent executions of NPU and PIM (Section~\ref{sec:scheduler}). 

\niparagraph{\circled{4} \neupims compiler framework.}
\neupims compiler framework is the frontend where the system admin provides the desired LLM and system specifications. 
The syntax of LLM specification largely resembles ONNX~\cite{onnx}. 
Upon receiving the specification, the compiler translates the model configuration into operations, each of which is represented as an intermediate representation (IR).
For the given IR, our compiler produces NPU and \neupims instruction binaries (i.e., Compute and MEM/PIM access instructions).
This process involves adjusting tile sizes and the sequence of instructions to align with the \neupims system specification.

\section{\neupims Architecture}
\label{sec:architecture}

\subsection{PIM Microarchitecture for Concurrent Execution}

\begin{figure}[t]
        \centering
        \includegraphics[width=1.0\linewidth]{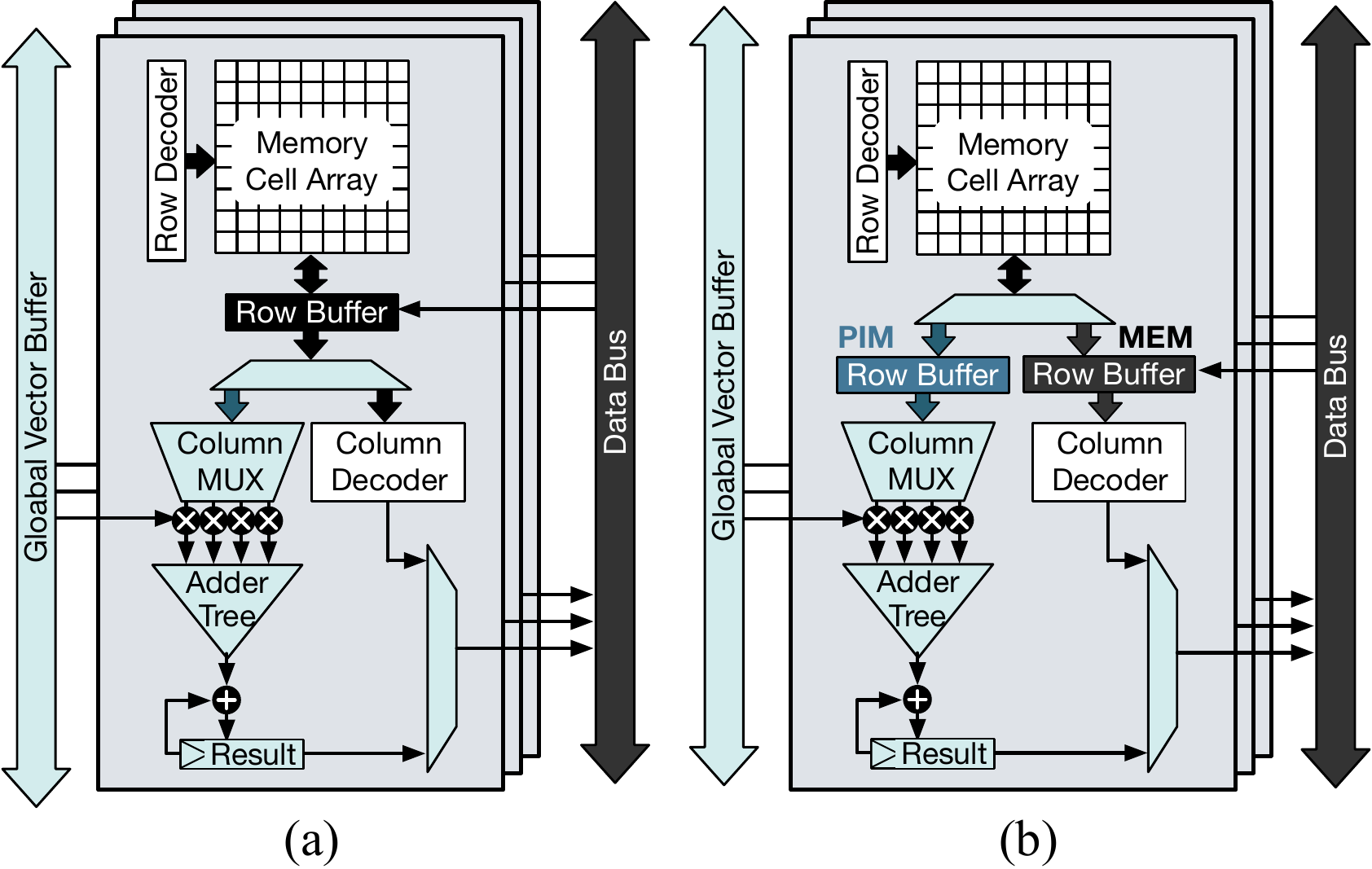}
        \vspace{-4ex}
        \caption{Microarchitecture of memory banks in (a) existing PIM accelerators with single row buffer banks, and (b) the proposed \neupims with dual row buffer banks.}
        \vspace{-2ex}
        \label{fig:newton-vs-neupims}
\end{figure}

\niparagraph{Single row buffer for PIM-based accelerator.}
Figure~\ref{fig:newton-vs-neupims}(a) depicts the high-level architecture of PIM-based GEMV accelerators that utilize banks equipped with a single row buffer.
For a GEMV, the vector operand is first located in the global buffer shared across all banks in a channel.
On the contrary, the rows of matrix operand are read from multiple banks simultaneously, exploiting bank-level parallelism, and located at their corresponding row buffers.
When the operands are ready, the parallel multipliers and adder tree perform a partial dot-product by reading the broadcast vector input from the global buffer and per-bank row buffers.

\niparagraph{Limitation of current PIM-based GEMV accelerators.}
Current PIM accelerators~\cite{newton, hbm_pim, aespa} operate in a ``blocked'' mode, preventing the simultaneous executions of NPU\footnote{
Henceforth in this paper, ``NPU'' refers exclusively to the device integrated within \neupims, setting it apart from any standalone NPUs.
} and PIM. 
This limitation arises primarily because the memory bank utilizes a single row buffer that serves two purposes: read/write memory operations and PIM acceleration specifically for GEMV.
These modes are managed sequentially, making it impossible to perform simultaneous executions. 
While this constraint does not significantly impact PIM system's performance for sole execution of GEMV, it becomes pertinent for LLM inferencing that requires both GEMM and GEMV operations.
Addressing this issue, our work aims to facilitate the parallel execution of both modes within the PIM framework. 
This advancement is expected to significantly enhance performance in LLM inferencing applications, unlocking the full potential of PIM beyond its current limitations.

\niparagraph{Extending PIM with dual row buffers.}
Figure~\ref{fig:newton-vs-neupims}(b) delineates the microarchitecture of \neupims bank. 
The memory banks of \neupims are equipped with \emph{dual row buffers}, namely MEM row buffer and PIM row buffer, which are associated with two independent data paths. 
Memory (MEM) row buffer is exclusively used for regular memory read/write accesses, whereas PIM row buffer is employed for GEMV operation. 
In designing \neupims, our design principle is to minimize the microarchitectural modification since the complication would impose significant area and power costs, lowering the practicality in the real-world system. 
Instead, we delegate the complications to the command interface and memory control mechanism, which will be elaborated below.

For prototyping and evaluating \neupims, we choose an industry-developed PIM accelerator designed for GEMV, Newton~\cite{newton}.
However, note that the proposed techniques in this work are not bounded to the Newton architecture, rather applicable to any GEMV accelerator that follows the standard DRAM microarchitecture and command interface. 

\subsection{Memory Command Interface} 
\niparagraph{Existing command interface for PIM-based GEMV.}
We develop the \neupims device using a PIM accelerator with a modified command interface on top of the existing DRAM standard interface.
There are four commands that collaboratively operate the PIM.
First, to process GEMV in PIM banks, \neupims must copy the operand vector to a global vector buffer shared by the banks within a channel. 
\neupims perform this operation using the PIM\_GWRITE command, which copies a specific row of a particular bank to the global vector buffer.
Similarly, it needs another command to activate the rows of the operand matrix into PIM row buffers across the banks.
For this, the host produces grouped activation commands, called ``PIM\_ACTIVATION'', which activate PIM row buffers of multiple banks simultaneously, usually for 4 banks at a time due to the power constraints (i.e., tFAW).
Once all the banks are activated, the host sends ``PIM\_DOTPRODUCT'' command  that performs parallelized dot-product computation, extracting massively larger in-memory bandwidth than host-memory bandwidth. 
Finally, the ``PIM\_RDRESULT'' command transfers the accumulated results to the host, ending a round of GEMV operation in PIM. 

\begin{figure}[t]
        \centering
        \includegraphics[width=0.9\linewidth]{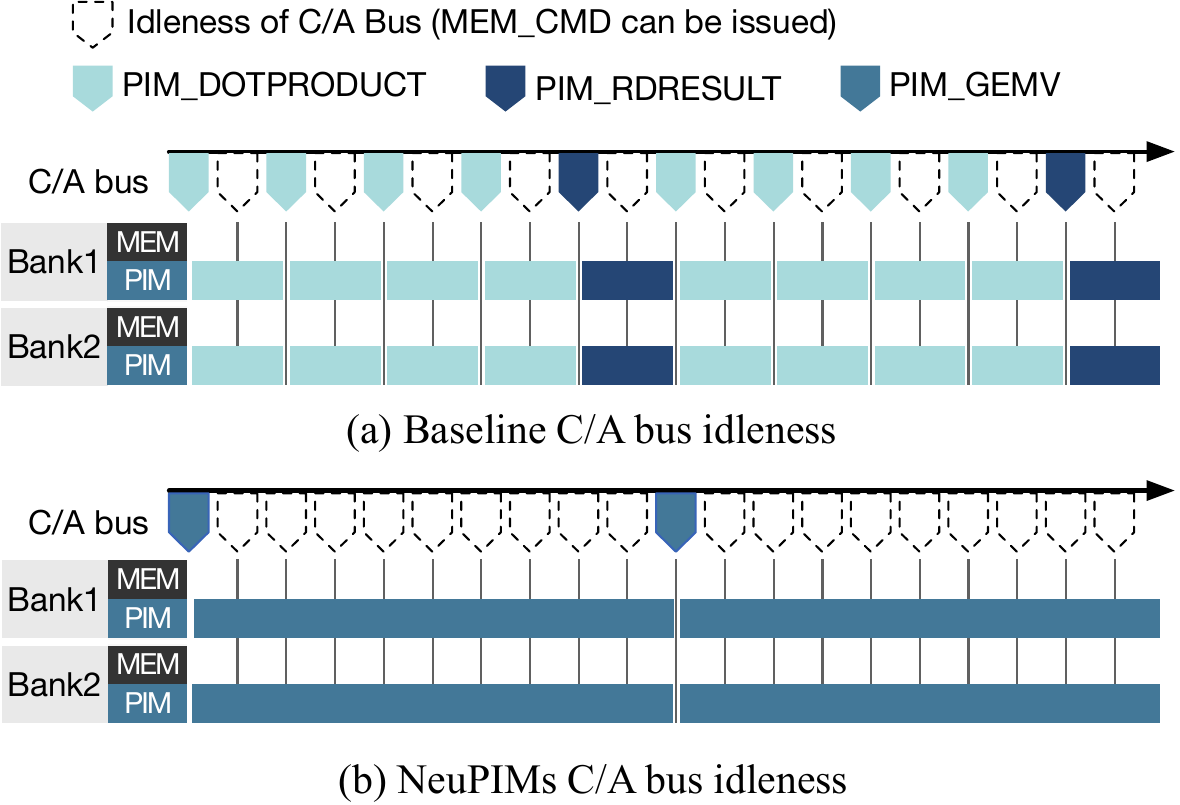}
        \vspace{-2ex}
        \caption{PIM command timing comparison.}
        \vspace{-2ex}
        \label{fig:body-pim-cmd-timing}
\end{figure}

\niparagraph{\neupims command interface.}
We modify the command interface of baseline PIM by augmenting three additional commands that support new capabilities for \neupims.
\begin{table}
\centering
\small 
\renewcommand{\arraystretch}{1.05}
\caption{List of \neupims commands.}
\footnotesize
\label{tab:neupims-commands}
\vspace{-2ex}
\begin{tabular}{c|c}

\hline
\textbf{Command} & \textbf{Description}                                                                                                                 \\ 
\hline
PIM\_HEADER                                                                           & \begin{tabular}[c]{@{}l@{}}Configure a GEMV operation \end{tabular} \\

PIM\_GEMV                                                                          & \begin{tabular}[c]{@{}l@{}} Perform $k$ dot-products and read the results   \\\end{tabular}                                                           \\
PIM\_PRECHARGE                                                                       & \begin{tabular}[c]{@{}l@{}}Precharge PIM row buffer\\\end{tabular}                                                          \\

\hline
\end{tabular}
\vspace{-2ex}
\end{table}

\begin{itemize}
\item \textbf{PIM\_HEADER: } 
This command allows varying dimensionalities of GEMV operation. 
Existing PIM accelerators have a rigid architecture supporting a GEMV with fixed dimensionalities, and therefore, the GEMV's execution latency is deterministic. 
This property allows the memory controller to deterministically schedule the PIM commands without violating the DRAM refresh intervals. 
However, as \neupims targets the GEMV operations of LLM's MHA layers, its dimensionality varies depending on the sequence length, and thus, the memory controller has no means to accurately calculate the latency, which may lead to DRAM refresh in the middle of PIM execution. 
To address this issue, \neupims allows the software to initialize a GEMV execution by sending the PIM\_HEADER command, which delivers the dimensionality information of scheduled GEMV operation. 
This way, the memory controller is able to estimate the end-to-end latency of GEMV operation and schedule its constituent commands without conflicting with the DRAM refresh.
\item \textbf{PIM\_GEMV:} 
The GEMV operation in existing PIM is controlled by a series of PIM\_DOTPRODUCT commands, and the result is read using the PIM\_RDRESULT commands, as depicted in Figure~\ref{fig:body-pim-cmd-timing}(a).
This fine-grained control approach naturally results in substantial command traffic in the memory C/A bus. 
While this is not an issue for PIM that operates in a blocked mode, it becomes a significant concern for \neupims that operate two functionalities in parallel.
PIM\_GEMV is a composite command that controls multiple dot-products simultaneously, and at the end, returns the result back to the host NPU. 
Figure~\ref{fig:body-pim-cmd-timing}(b) shows an example timeline that shows the reduced command traffic by PIM\_GEMV.
The number of dot-products, $k$, is given as an argument for the command.

\item \textbf{PIM\_PRECHARGE:} 
As the \neupims banks have dual row buffers, there is a need for an additional command that specifically precharges the PIM row buffers once the GEMV is completed. 
PIM\_PRECHARGE is the same as the regular PRECHARGE command except that it triggers precharge of the PIM row buffer. 
\end{itemize}

\subsection{Memory Controller}
For \neupims, we have multiple channels, each of which has multiple PIM banks. 
The LLM requests are assigned to one of these channels, with the MHA layer execution for each request being distributed across the channel's multiple PIM banks.
The memory controllers located in their respective PIM channels are equipped with their individual PIM command queues.
PIM commands are broadcast to all banks in the corresponding channel.

\niparagraph{Interleaved scheduling of memory read/write and PIM commands.}
A challenge in the implementation of \neupims memory controller is to interleave the memory read/write commands and PIM commands efficiently so that the command/address bus bandwidth does not become a performance bottleneck.
\neupims prioritize PIM commands over memory read/write commands, since the issuing delay of PIM commands is greater than that of memory commands, so the C/A bus bandwidth used to issue PIM commands is relatively small, allowing both commands to be issued without significant performance degradation.

\section{\neupims Scheduling}
\label{sec:scheduler}

The integration of dual row buffers enables \neupims to handle NPU's memory accesses and PIM commands simultaneously.
This section describes computation overlapping opportunities for multi-head attention (MHA) layer execution and a novel request scheduling technique for batched LLM inferencing, dubbed \textit{sub-batch interleaving}.

\subsection{Overlapping Opportunities in MHA Layer}
\begin{figure}[t]
\centering
\includegraphics[width=1.0\linewidth]{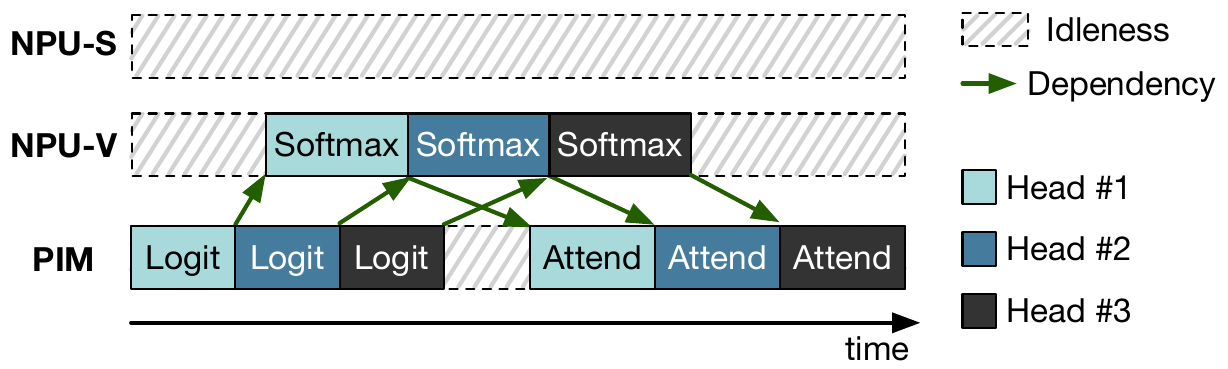}
\vspace{-3ex}
\caption{Overlapping opportunities of multi-head attention layers. NPU-S: systolic arrays; NPU-V: vector units.}
\vspace{-1ex}
\label{fig:overlap-mha}
\end{figure}

\begin{figure}[t]
        \centering
        \includegraphics[width=1.0\linewidth]{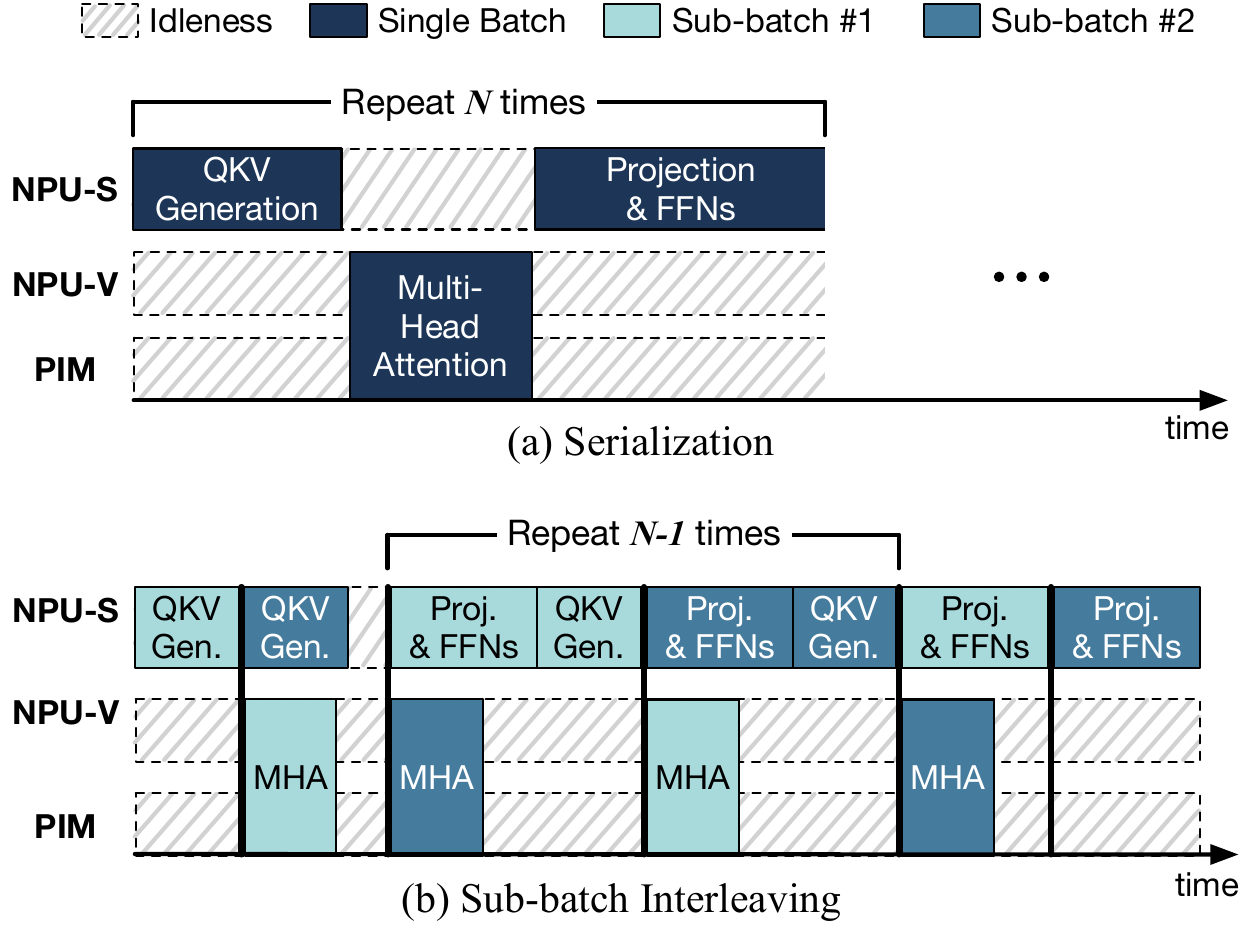}
        \vspace{-3ex}
        \caption{Example execution timelines of LLM decoder blocks: (a) Serialized execution, (b) Sub-batch interleaving. NPU-S: systolic arrays; NPU-V: vector units; \(N\): the number of decoder blocks.}
        \vspace{-3ex}
        \label{fig:sub-batch-interleaving}
\end{figure}

Figure~\ref{fig:overlap-mha} illustrates the overlapping of (1) logit and attend operations on PIM-side, and (2) softmax operations on NPU-side in \neupims.
As operations of multi-head attention can be decomposed to a head granularity, even na\"ive NPU-PIM integrated architecture should have an opportunity to harness both resources to overlap operations.
However, it cannot take advantage of this opportunity because operational results cannot be transferred between PIM units and vector units through PIM channels.
In contrast, \neupims, by employing dual row buffers, can concurrently harness both NPU and PIM.
This configuration allows vector units within \neupims to store partial logit (softmax) values without having to wait for the completion of the entire logit operations (GEMV) on the PIM, reducing the underutilization of integrated units.


It is worthwhile to note that this overlapping is only possible because there is head-level parallelism, which is only available within the multi-head attention layer.
Further, the overlapping opportunity exists only between PIM and vector units of NPU, resulting in the NPU's systolic arrays largely remaining unused during the execution of the MHA layer.

\subsection{Sub-batch Interleaving}
\label{subsec:sub-batch}
\niparagraph{Limitation of serialized executions.}
Figure~\ref{fig:sub-batch-interleaving}(a) illustrates an example execution timeline of the decoder block operations, running on a na\"ive NPU-PIM integrated device.
In particular, the figure depicts the algorithmic dependencies among QKV generation, multi-head attention, and projection \& FFNs, within the batch. 
Due to the dependencies, they must be executed serially, inevitably leading to low utilization on both NPU and PIM.  

\niparagraph{Interleaving the two sub-batches.}
To tackle this challenge, we propose \textit{sub-batch interleaving} that partitions one large batch into two sub-batches and alternate them to improve resource utilization.    
Figure~\ref{fig:sub-batch-interleaving}(b) delineates the sub-batch interleaving technique that allows the simultaneous executions of PIM-friendly and NPU-friendly operations within the sub-batches on the PIM and NPU, respectively, significantly improving the utilization of both NPU and PIM.

\niparagraph{Comparative analysis on the execution timelines.}
Let $N$ denote the number of decoder blocks to be executed on a single \neupims device.
Figure~\ref{fig:sub-batch-interleaving}(a) shows that without the use of sub-batch interleaving, the operators in each decoder block would be executed sequentially, leading to a total execution time equal to $N$ times the per decoder-block execution time (i.e., QKV generation on NPU-S + MHA on PIM \& NPU-V + Projection \& FFNs on NPU-S). 
However, as depicted in Figure~\ref{fig:sub-batch-interleaving}(b), sub-batch interleaving effectively hides the execution times of MHA layer in the NPU-S execution times, resulting in the total execution time equal to  ($N$ - 1) times the per-sub-batch partial execution time (i.e., Projection, FFNs, and QKV generation on NPU-S) plus a single decoder-block execution time divided at the start and end.
While the interleaving occurs, the NPU and PIM utilization is improved as their executions are effectively overlapped.
Our empirical study demonstrates that the \neupims execution time in the interleaved period is mostly bounded by the NPU execution time running GEMM operations, hiding the PIM execution time for MHA layer execution.

\niparagraph{Challenges.}
For optimal exploitation of parallelism inherent in sub-batch interleaving, \neupims must consider two key aspects:
First, \neupims requires a strategic balancing of the execution time for each sub-batch, particularly focusing on the multi-head attention.
Since the latency of multi-head attention is determined by the channel processing the longest sequence, we must implement load balancing across the channels, ensuring an equitable distribution of token lengths.
This challenge is addressed through a channel load balancing algorithm (Section~\ref{subsec:channel-load-balancing}).
Second, \neupims must ensure similar execution times for both sub-batches for efficient interleaved execution.
The duration of each stage within the interleaving is bound by the processing time of more time-consuming sub-batch.
For that, \neupims introduces a sub-batch partitioning algorithm (Section~\ref{subsec:sub-batch-partition}).

\SetKwInOut{Input}{input}
\SetKwInOut{Output}{output}
\SetKwInput{KwData}{Parameter}
\SetKwComment{Comment}{\textsf{//} }{}
\SetKwComment{CommentTwo}{/* }{~*/}

\begin{algorithm}[tb]
\footnotesize
\caption{MHA Latency Estimation}
\label{alg:estimate-mha}
\LinesNumbered

\KwIn{$
\hspace{0.2em}\quad\quad\ {seq\_len}:\   \small\textit{Sequence length of the request} 
$}
\KwData{$
\hspace*{0.2em}{E}:\     \small\textit{Model embedding size}\newline
\hspace*{0.2em}{L}_{tile}:\   \small\textit{GEMV latency for one PIM tile} \newline 
\hspace*{0.2em}{L}_{GWRITE}:\  \small\textit{GWRITE latency for global buffer} \newline 
\hspace*{0.2em}{E}:\  \small\textit{Model embedding size}\newline
\hspace*{0.2em}{P}_{DRAM}:\  \small\textit{DRAM page size}\newline
\hspace*{0.2em}{B}_{chnl}:\  \small\textit{PIM banks per channel}\newline
\hspace*{0.2em}{N}_{head}:\  \small\textit{Number of heads}
$}
\KwOut{$
\hspace{0.2em} \quad\ \ {L}_{MHA}:\ \small\textit{Estimated latency for MHA}
$}

\vspace{1ex}
$L_{MHA}\ \leftarrow\ 0$\; \CommentTwo{\textit{\textsf{\footnotesize{\textbf{GEMV latency for \ensuremath{Key^T \times\ Query}}}}}}
$N_{tiles}\ \leftarrow\ ({seq\_len}\ /\ {B}_{chnl})\ *\ (E\ /\ P_{DRAM})$\;
${L}_{MHA}\ +=\ L_{GWRITE}\ *\ (E\ /\ P_{DRAM})$\;
${L}_{MHA}\ +=\ L_{tile}\ *\ N_{tiles}$\;
\vspace{1ex}
\CommentTwo{\textit{\textsf{\footnotesize{\textbf{GEMV latency for \ensuremath{Logits\ \times\ Value}}}}}}
$N_{tiles}\ \leftarrow\ ((E/N_{head})/B_{chnl})*((seq\_len/P_{DRAM})*N_{head})$\;
${L}_{MHA}\ +=\ L_{GWRITE}\ *\ ((seq\_len\ /\ P_{DRAM})\ *\ N_{head})$\;
${L}_{MHA}\ +=\ L_{tile}\ *\ N_{tiles}$\;
\Return{$L_{MHA}$}
\end{algorithm}
\vspace{-1ex}

\subsection{Multi-Head Attention Latency Estimation}
\label{subsec:estimate-mha}
The latency of operations running in the NPU is largely dependent on the batch size of the inference. 
To apply optimization techniques for multi-head attention, we estimate the execution time of its operations by considering the key-value mapping to the PIM memory layout. 
Since the vector for GEMV is shared across the banks, the matrix for GEMV is interleaved row-wise to each banks.
Consequently, key caches at the same row and column share the same layer and head index, with differing sequence indices. 
Conversely, value caches at the same row and column share the same layer, head, and sequence index, interleaving each head embedding into banks. 
Algorithm~\ref{alg:estimate-mha} takes this mapping into account to estimate the execution time of the multi-head attention latency.

\SetKwInOut{Input}{input}
\SetKwInOut{Output}{output}
\SetKwInput{KwData}{Parameter}
\SetKwComment{Comment}{\textsf{//} }{}
\SetKwComment{CommentTwo}{/* }{~*/}

\SetKwFunction{GetLoad}{\textnormal{GetLoad}}
\SetKwFunction{Min}{\textnormal{Min}}
\SetKwFunction{append}{\textnormal{append}}
\SetKwFunction{MHAest}{\textnormal{MHALatencyEstimation}}
\SetKwFunction{Index}{\textnormal{index( )}}

\begin{algorithm}[tb]
\footnotesize
\caption{Greedy Min-Load Bin Packing}
\label{alg:channel-load-balancing}
\LinesNumbered
\KwIn{$
    \hspace{0.25em}{L_{req}}:\ \small\textit{A list for sqeunce length of new requests }\newline 
    \hspace{0.25em}\ \ \ {L_{chnl}}:\ \small\textit{A list for request allocation of channels }
$}

\vspace{1ex}
\CommentTwo{\textit{\textsf{\small{Calculate each channel's total load by applying \newline MHA latency estimation to each allocated request }}}}
$L_{load}\ \leftarrow\ []$\;
\ForEach{$chnl$ in \ $L_{chnl}$}{
    $Sum_{load}\ \leftarrow\ 0$\;
    \ForEach{$req$ in \ $chnl$}{
        $Sum_{load}\ +=\ \MHAest{req}$
    }
    $L_{load}.\append{\ensuremath{Sum_{load}}}$\;
}
\vspace{1ex}
\CommentTwo{\textit{\textsf{\small{Allocate each request by greedy algorithm}}}}
\ForEach{$new\_req$ in \ $L_{req}$}{
    $min\_index\ =\ {\Min{\ensuremath{L_{load}}}.\Index}$\;
    $L_{chnl}[min\_index].{\append{new\_req}}$\;
    $load_{req}\ =\ \MHAest{new\_req}$\;
    $L_{load}[min\_index]\ +=\ {\ensuremath{load_{req}}}$\;
}
\Return{$L_{chnl}$}
\end{algorithm}
\vspace{-1ex}

\subsection{Greedy Min-Load Bin Packing Algorithm.}
\label{subsec:channel-load-balancing}

To minimize the execution time discrepancy between the most
congested channel and the load-free channel, we developed
the channel load balancing algorithm. Algorithm~\ref{alg:channel-load-balancing} leverages
the aforementioned multi-head attention latency estimation
to batch requests. It initially sorts the batch of requests in
decreasing order of input token length. Then, \neupims
places a request with the longest token length in the channel
with minimal load. At each iteration, it updates the estimated
latency, taking into account the newly appended request.

\neupims allocate LLM inference requests to one of the PIM channels.
A PIM channel constitutes multiple PIM banks, each of which partially executes the MHA layers of the assigned requests.
Thus, it is crucial to minimize the execution time discrepancy between the most congested channel and the load-free channel for load-balancing purposes. 
To this end, we develop \emph{greedy min-load bin packing algorithm}, as presented in Algorithm~\ref{alg:channel-load-balancing}.
This algorithm leverages the aforementioned multi-head attention latency estimation to batch requests. 
As implied by its name, the algorithm greedily allocates requests starting from the longest sequence length to the least loaded PIM channel sequentially.
Therefore, it initially sorts the batch of requests in decreasing order of input token length. 
Then, \neupims places a request with the longest token length in the channel with minimal load. 
At each iteration, it updates the estimated latency, taking into account the newly appended request.

\SetKwInOut{Input}{input}
\SetKwInOut{Output}{output}
\SetKwInput{KwData}{Parameter}
\SetKwComment{Comment}{\textsf{//} }{}
\SetKwFunction{Size}{\textnormal{Size}}
\SetKwFunction{Ceil}{\textnormal{\textbf{ceil}}}
\SetKwFunction{Floor}{\textnormal{\textbf{floor}}}
\SetKwFunction{append}{\textnormal{append}}

\begin{algorithm}[tb]
\footnotesize
\caption{Sub-Batch Partitioning }
\label{alg:sub-batch-partition}
\LinesNumbered

\KwIn{$
    \hspace{0.25em}\ \ \ {L_{req}}:\ \small\textit{A list of active request set in each channel}
$}

\KwOut{$
    \hspace{0.25em}{SB_1,\ SB_2}:\ \small\textit{Sub-batchs for interleaving}
$}
\vspace{1ex}

$turn\ \leftarrow\ True$;

$SB_1,\ SB_2\ \leftarrow\ []$\;
\ForEach{$req_{chnl}$ in $L_{req}$}{
        $bsize\ \leftarrow\ \Size{\ensuremath{req_{chnl}}}\ /\ 2$\;

    \If{$\Size{\ensuremath{req_{chnl}}}\ \%\ 2\ !=\ 0$}{
            $bsize\ =\ turn\ ?\ \Ceil{\ensuremath{bsize}}\ :\
            \Floor{\ensuremath{bsize}}$\;

            $turn\ =\ !turn$\;
    }

        $bsize\ =\ \mathbf{int}({\ensuremath{bsize}})$;

    \vspace{1ex}
        $SB_1.\append{\ensuremath{req_{chnl}[:\ bsize]}}$\;
        $SB_2.\append{\ensuremath{req_{chnl}[bsize\ :]}}$\;

         
}
\Return{$SB_1,\ SB_2$}

\end{algorithm}

\subsection{Sub-batch Partitioning Algorithm}
\label{subsec:sub-batch-partition}
Given the dependency of NPU-friendly operations on the batch size of inference, it is essential to maintain a balanced size between the two sub-batches. 
As outlined in Algorithm~\ref{alg:sub-batch-partition}, the approach involves dividing the requests for each channel into halves and appending each half to one of the sub-batches.

\section{Scaling \neupims System}
\label{sec:scale}

Model parallelism, which involves partitioning model parameters across multiple \neupims devices for parallel processing, is essential for inference tasks of LLM, due to the limited memory capacity of \neupims devices.
In this section, we will discuss the applicability of \neupims systems to pipeline parallelism and tensor parallelism, two widely used model parallelism techniques in deep learning libraries~\cite{megatronlm, tensorrt, tensorrt-llm} for LLM inference.
Note that, these model parallelism strategies are not the contribution of this work, but this section highlights the adaptability of such techniques to \neupims system.

\subsection{Pipeline Parallelism of \neupims System}
Pipeline parallelism involves dividing the model layer-wise, so that several layers of model are placed on a single \neupims device. 
To facilitate parallel processing, the batch is divided into micro-batches corresponding to the pipeline depth, and each device processes them in a pipelined manner. 
This approach can also be applied to \neupims systems in a similar manner.
When applying pipeline parallelism to the \neupims systems, the number of decoder blocks executed on a single \neupims device decreases proportionally. 
As mentioned in Section~\ref{subsec:sub-batch}, the reduced decoder blocks co-located in one \neupims device would lower the achievable performance benefits.
Furthermore, exploiting pipeline parallelism leads to smaller batch size.
Using the sub-batch interleaving technique further reduces batch size, which would lead to under-utilization of the NPU systolic arrays. 
\subsection{Tensor Parallelism of \neupims System}
Tensor parallelism is a parallelization approach that splits the model tensors into multiple shards and distributes them across devices.
Multiple devices execute on their respective model shards in parallel, the results of which are aggregated at the end of the step. 
This aggregation requires communication between devices.
Although the sub-batch interleaving technique increases communication frequency twofold, the total communication traffic remains unchanged compared to the non-partitioned batch, resulting in only a modest overhead from communication.
Moreover, the sub-batch that completes first among the two sub-batches can engage in communication, while the other sub-batch performs computations. 
This further reduces the communication latency, mitigating the increased communication overhead.
Inspired by the aforementioned insights, we prioritize the exploitation of tensor parallelism over pipeline parallelism and start employing pipeline parallelism when the model size is too large to exclusively leverage tensor parallelism.
\section{Evaluation}
\label{sec:method}
\subsection{Methodology}

\niparagraph{Baseline.}
For evaluation, we compare three baseline systems along with our \neupims system, namely GPU-only, NPU-only, and NPU+PIM.
\begin{itemize}[labelindent=0.1em,nolistsep,leftmargin=1em]
    \item \textbf{\textsf{GPU-only}:}
    \textsf{GPU-only} is a real GPU system. We conduct experiments with an A100 40GB GPU, and we compile the LLM batched inference workload with PyTorch~\cite{pytorch}.
    
    \item \textbf{NPU-only:}
    NPU-only represents existing NPU accelerators such as TPU, without any PIM capabilities. 
    We assume that this baseline has the equivalent memory bandwidth with other alternatives for fairness. 
    Moreover, a NPU is equipped with not only systolic array but also GPU-like vector processing units to support \emph{non}-GEMM operators.  
    \item \textbf{NPU+PIM:}
    NPU+PIM is also a PIM-enabled NPU baseline, which integrates existing PIM-based GEMV accelerators~\cite{newton} with the off-the-shelf NPU accelerator. 
    We merely map the GEMV operations in MHA layers to the PIM, while all the others are computed at the NPU side.
    We allocate the requests to PIM channels in a round-robin manner.
\end{itemize}

\niparagraph{Cycle-level simulation.}
We build the \neupims simulator on two open-source cycle-accurate simulators, ONNXim~\cite{onnxim} and DRAMsim3~\cite{dramsim3}. 
We link the two simulators by modifying the memory interface of ONNXim and offloading the memory accesses to the PIM simulator based on DRAMsim3.

\begin{table}[]
\caption{\neupims hardware specification.}
\footnotesize
\vspace{-2ex}
\label{tab:hardware-spec}
\begin{tabular}{cccc}
\hline
\multicolumn{2}{c|}{\textbf{NPU Configuration}}                                                         & \multicolumn{2}{c}{\textbf{HBM Organization}}                             \\ \hline
\multicolumn{1}{c|}{Systolic Arrays/Chip}                & \multicolumn{1}{c|}{8}                       & \multicolumn{1}{c|}{Banks/Bankgroup}                 & 4                  \\
\multicolumn{1}{c|}{Vector Units/Chip}                   & \multicolumn{1}{c|}{8}                       & \multicolumn{1}{c|}{Banks/Channel}                   & 32                 \\
\multicolumn{1}{c|}{HBM Channels/Chip}                   & \multicolumn{1}{c|}{32}                      & \multicolumn{1}{c|}{Frequency}                           & 1GHz               \\
\multicolumn{1}{c|}{Systolic Array Size}                 & \multicolumn{1}{c|}{128 x 128}               & \multicolumn{1}{c|}{Capacity/Channel}                & 1GB                \\
\multicolumn{1}{c|}{Vector Unit Size}                    & \multicolumn{1}{c|}{128 x 1}               & \multicolumn{1}{c|}{Page Size}                       & 1KB                \\ \hline
\multicolumn{4}{c}{\textbf{HBM Timing Parameter}}                                                                                                                                   \\ \hline
\multicolumn{4}{c}{\begin{tabular}[c]{@{}c@{}}tRP = 14, tRCD = 14, tRAS = 34, tRRD\_L = 6, tWR = 16, \\ tCCD\_S = 1, tCCD\_L = 2, tREFI = 3900, tRFC = 260, tFAW = 30\end{tabular}} \\ \hline
\end{tabular}
\vspace{-2ex}
\end{table}
\niparagraph{Hardware specifications.}
We prototype \neupims accelerator using a set of hardware specifications, which are listed in Table~\ref{tab:hardware-spec}.
Our \neupims accelerator prototype is a multi-chiplet design containing 8 systolic arrays, each integrated with a SIMD vector unit that serves activation operations.
Each memory channel controls 32 PIM banks, which offer in aggregate 1GB memory capacity. 
Note that while we choose this set of specifications for prototyping purposes, the \neupims architecture is orthogonal to these choices, allowing varying configurations depending on the model size and data-specific properties (e.g., sequence lengths). 

\begin{figure*}[t]
        \centering
        \includegraphics[width=1\linewidth]{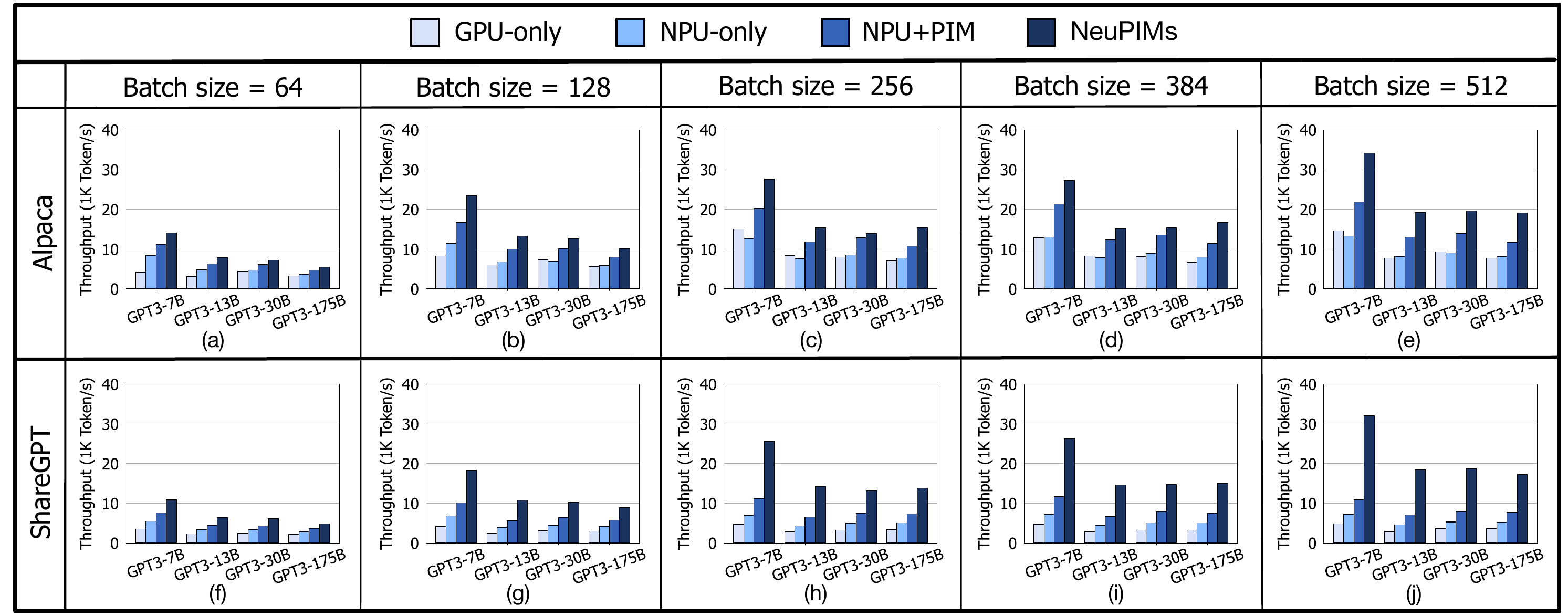}
        \vspace{-2ex}
        \caption{Throughput comparison results of GPU-only, NPU-only, NPU+PIM, and \neupims. We use the two datasets, (a) Alpaca and (b) ShareGPT, using the batch sizes including 64, 128, 256, 384, and 512.}
        \vspace{-2ex}
        \label{fig:eval-end-to-end}
\end{figure*}

\begin{table}[]
\renewcommand{\arraystretch}{1.1}
\centering
\caption{The evaluated LLM configurations.}
\vspace{-2ex}
\label{tab:model-config}
\footnotesize
\begin{tabular}{c|ccc|cc}
\hline
\textbf{Model}     
& \textbf{$\#$ Layers}
& \textbf{$\#$ Heads}
& \textbf{d$_{model}$}
& \textbf{\# TP}
& \textbf{\# PP}
\\ \hline

GPT3-7B & 32 & 32 & 4096 & 4 & 1 \\
GPT3-13B  & 40 & 40 & 5120 & 4 & 1 \\
GPT3-30B  & 48 & 56 & 7168 & 4 & 2 \\
GPT3-175B & 96 & 96 & 12288 & 8 & 4 \\ 
\hline
\end{tabular}
\vspace{-2ex}
\end{table}

\niparagraph{LLM models.}
We use four variants of GPT3, a state-of-the-art LLM developed by OpenAI as described in Table~\ref{tab:model-config}.
While our experiments focus on GPT-3 model variants, \neupims can host any decoder-based generation models, offering generality and wide applicability. 

\niparagraph{Datasets.}
We use two real-world LLM inferencing datasets, ShareGPT~\cite{sharegpt} and Alpaca~\cite{alpaca}. 
ShareGPT dataset is a set of conversations scraped from the real-world user log of ChatGPT~\cite{openai-chatgpt}. 
Alpaca dataset is an instruction dataset generated by OpenAI's text-davinci-003 engine. 
The two datasets have distributions for input and output sequences. 
ShareGPT has an average input token length of 80 and an output of 296, while Alpaca has shorter sequence lengths of 12 and 56 for input and output, respectively.

\niparagraph{Workload.}
As running experiments for inference serving scenarios with a cycle-accurate simulator is infeasible, we develop an alternative methodology to synthesize workloads for system-level evaluation. 
To define the search space for workloads, we consider various hyperparameters, including model types, batch sizes, and tensor/pipeline parallelism.
For each permutation of these hyperparameters, we simulate the inference serving for a fixed amount of time, randomly picking sequence lengths from the datasets. 
This way, we can warm up the inference batch in a way that the batch is filled with requests having various sequence lengths. 
We sample 10 different batches and use them to measure the throughputs of different hyperparameter combinations.
\subsection{Results}
\label{sec:eval}
\niparagraph{Throughput.}
Figure~\ref{fig:eval-end-to-end} reports the throughput comparison results between the three baselines and \neupims. 
The GPU-only and NPU-only baseline systems show marginal differences since they both execute end-to-end decoder block operations without PIM, including bandwidth-bound multi-head attention.
Simply integrating PIM with the NPU, i.e., NPU+PIM already offers, on average 1.5$\times$ throughput improvement compared to the NPU-only baseline because the bandwidth-bound GEMV operations in MHA are all offloaded to the PIM.
However, \neupims consistently surpasses the NPU+PIM baseline, and offers \emph{additional} throughput improvements over the NPU+PIM baseline across all the models and datasets, ranging from 13\% to 3$\times$.
These trends are observed consistently for both datasets, with larger gains observed for ShareGPT, given its longer input/output sequences, offering increased acceleration opportunities for PIMs.
Furthermore, as the batch size increases from 64 to 512, the throughput gains exhibit substantial growth. 
This is because the \neupims system effectively shifts the bottleneck from bandwidth to compute towards the NPU, thus allowing \neupims to extract higher performance from batched computation as the batch size grows. 
Note that \neupims achieves significant throughput improvements using the same memory capacity, which demonstrates its cost-effectiveness, especially in a datacenter-scale inference serving scenario where larger batch sizes are advantageous.  

\begin{table}[]
\centering
\caption{Average resource utilization of NPU/PIM compute resource and memory bandwidth utilization.}
\label{tab:util}
\small
\vspace{-2ex}
\begin{tabular}{c|c|c|c}
\hline
          & \textbf{NPU-only}          & \textbf{NPU+PIM}          & \textbf{\textsc{NeuPIMS}}         \\ \hline
NPU       & 12.3\%         & 28.0\%             & \textbf{64.9}\%   \\
PIM       & -               & 17.0\%             & \textbf{26.4\%}   \\
Bandwidth & 67.6\%         & 27.4\%             & \textbf{85.4\%}   \\
\hline
\end{tabular}
\end{table}
%
\niparagraph{Utilization.}
The main source of large throughput gain is the improved resource utilization across the board. 
Table~\ref{tab:util} compares the average utilization for three major resources in the baselines and \neupims, when using GPT3-30B, the batch size of 256, and the ShareGPT dataset.  
As NPU+PIM offloads the bandwidth-bound MHA operations to PIM, it increases NPU utilization by 28.0\%.
However, NPU+PIM still suffers from temporal blocking due to the GEMM-GEMV dependencies in LLM inferencing. 
\neupims overcomes this limitation, achieving 64.9\% and 26.4\% utilization on NPU and PIM, respectively, through the concurrent NPU+PIM execution capability.
We observe similar trends from the other system configurations, which demonstrate the effectiveness of the proposed technique in enabling concurrent NPU+PIM execution by \neupims.

\begin{figure}[t]
        \centering
        \includegraphics[width=0.95\linewidth]{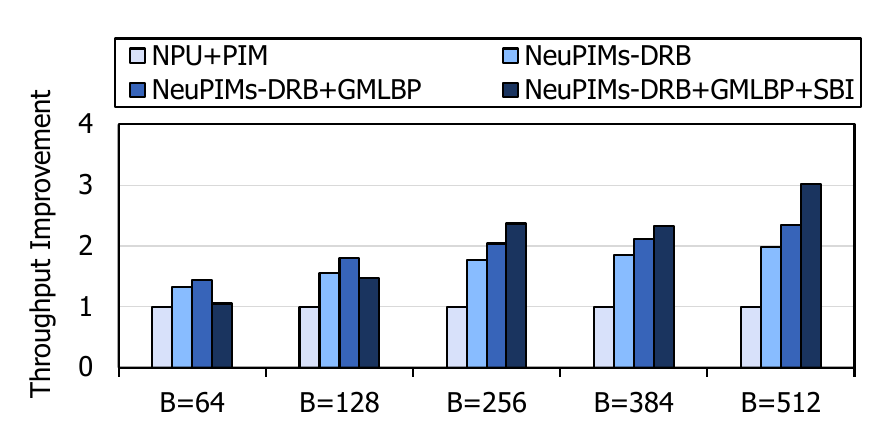}
        \vspace{-2ex}
        \caption{We use the GPT3-7B model and ShareGPT dataset for this experiment. DRB: Dual Row Buffers; GMLBP: Greedy Min-Load Bin Packing algorithm; SBI: Sub-Batch Interleaving.}
        \vspace{-2ex}
        \label{fig:batch-size-sensitivity}
\end{figure}

\niparagraph{Ablation study.}
Using NPU+PIM as the baseline, we augment the proposed three techniques and observe the performance behaviors, shown in Figure~\ref{fig:batch-size-sensitivity}. 
For all batch sizes, the dual row buffers offer 69.7\% throughput improvement on average, which has the largest impact on the performance as it enables concurrent NPU+PIM execution without significant overhead. 
For the channel loading balancing, our greedy min-load bin packing algorithm also always offers performance benefits by evenly distributing the requests to the available channels.
In contrast to the previous two techniques, the sub-batch interleaving technique does not always yield gains. 
For small batch sizes, partitioning the batch into two may cause underutilization in a NPU systolic array, leading to inefficiency, and the penalty from pipelining could outweigh the benefits.
However, when the batch size is equal to or larger than 256, \neupims achieve the highest throughput, suggesting that batched inference with a large batch size is preferable for the \neupims-based system.

\begin{figure}[]
        \centering
        \includegraphics[width=0.95\linewidth]{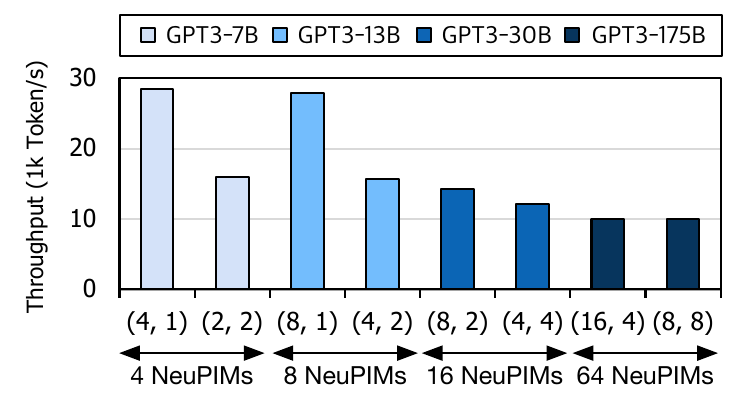}
        \vspace{-1ex}
        \caption{Throughput of multi-\neupims system as the parallelization schemes change. We employ four combinations of tensor parallelism (TP) and pipeline parallelism (PP), represented as (TP, PP).} 
        \vspace{-2ex}
        \label{fig:system-scale-sensitivity}
\end{figure}

\niparagraph{Implication of parallelization schemes.}
As the LLM size increases, the \neupims system must scale the number of devices to harness tensor and pipeline parallelisms.
Figure~\ref{fig:system-scale-sensitivity} analyzes the implications of such parallelization schemes on the system throughput. 
For the experiment, we fix the total number of requests to 256, while the batch size per device varies depending on the parallelization scheme. 
The results highlight the preference for exploiting tensor parallelism over the pipeline counterpart, as it maintains a large batch size, resulting in better efficiency at the NPU. 
We observe this trend consistently for all model variants, while the overall throughput decreases since the per-device batch size becomes small, due to low NPU utilization. 

\niparagraph{Area overhead.}
The main source of area overhead in the \neupims architecture is the dual row buffer. 
We use CACTI 7.0~\cite{cacti-v7} with 22nm technology to measure the area overhead by doubling the row buffer resource usage in the tool configuration.
We observe 3.11\% area overhead, which is marginal considering the significant performance boost provided the microarchitectural addition. 

\begin{table}[t]
\caption{\neupims power overhead.}
\footnotesize
\label{tab:power}
\vspace{-2ex}
\begin{tabular}{l|ll}
\hline
\multicolumn{1}{c|}{\textbf{Baseline}} & \multicolumn{2}{c}{\textbf{Average Power}}    \\ \hline
NPU-only                               & \multicolumn{1}{l|}{HBM (non-PIM)}         & 364.1 mW \\
\neupims                                & \multicolumn{1}{l|}{Dual row buffered PIM} & 634.8 mW \\ \hline
\end{tabular}
\vspace{-3ex}
\end{table}

\niparagraph{Power overhead.}
Compared to the NPU-only counterpart, \neupims requires higher power in memory because it operates NPU and PIM concurrently. 
We examine this power implication by measuring it using Micron's DRAM power model~\cite{micron-power} provided by DRAMsim3~\cite{dramsim3}. 
We assume all-bank computation command incurs 4$\times$ higher power than read command~\cite{newton}. 
Moreover, the ``additional'' row buffer requires DRAM to consume more background power to hold each row buffer status. 
We measure the total power overhead by aggregating all these factors together.  
Table~\ref{tab:power} compares the average power of \neupims and NPU-only system where NPUs are equipped with vanilla HBMs. 
\neupims exhibit 1.8$\times$ higher power consumption, offering 2.4$\times$ speedup, which can be translated into 25\% energy reduction.

\begin{figure}[t]
        \centering
        \includegraphics[width=1.0\linewidth]{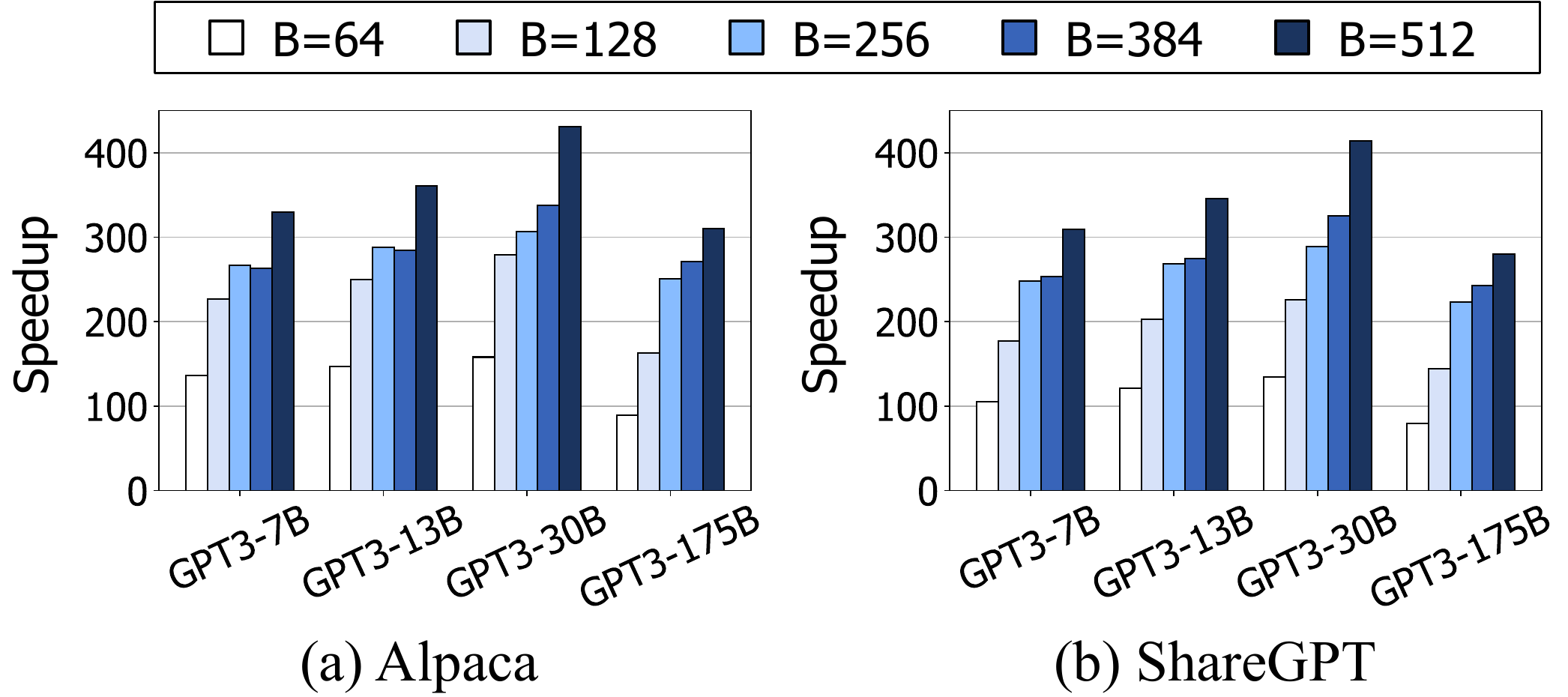}
        \vspace{-4ex}
        \caption{Speedup of \neupims over TransPIM~\cite{hpca22-transpim}.}
        \vspace{-4ex}
        \label{fig:speedup-over-transpim}
\end{figure}

\niparagraph{Comparison with TransPIM.}
TransPIM~\cite{hpca22-transpim} is a standalone PIM-only solution that operates the entirety of transformer operators within PIM devices.
As there is no open-source simulator for TransPIM, we develop our own TransPIM simulator based on DRAMsim3~\cite{dramsim3}.
We align the memory specifications of TransPIM, such as HBM timing parameters and capacity, with those used for \neupims and the NPU+PIM baseline.
Figure~\ref{fig:speedup-over-transpim} reports the speedup of \neupims over TransPIM~\cite{hpca22-transpim}
\neupims shows an average 228$\times$ higher throughput than TransPIM.
The significant performance gap is attributed to the effectiveness of GEMM computation executed on the NPU in the case of \neupims, as opposed to PIM in TransPIM.
That is, TransPIM specifically targets single-batch transformer model inference, making it unsuitable for batched inference. 
Additionally, we observe that the token-based dataflow and ring-broadcast mechanism proposed in TransPIM are designed to target encoder block operations, making them inefficient for decoder-based LLM inference.
Overall, \neupims consistently deliver superior performance compared to TransPIM, achieving speedups ranging from 79$\times$ to 431$\times$.
\section{Discussion}
\label{sec:discuss}

\niparagraph{Model training.}
As training has fixed-length sequences for both input and output, it entirely entails GEMMs, not requiring any GEMVs for its computations. PIM targets accelerating bandwidth-bounded GEMV operations, delivering significantly poorer performance for GEMMs. Therefore, while a \neupims system can be used for training, its efficiency is limited.

\niparagraph{Integration with production software stack.}
As described in Section~\ref{sec:neupims}, the \neupims compiler framework employs a similar interface with the modern machine learning libraries such as ONNX, PyTorch, and JAX.
Therefore, the integration of \neupims solution with the existing production software stack would require one to write a translator, which can convert the ONNX-, PyTorch-, and Jax-defined model representations into our LLM specification.
With the translator being developed, the rest of \neupims system would remain the same since it is already a holistic system stack that constitutes an inference serving scheduler, operator compilers for NPU and PIM, and an inference execution runtime.
\section{Related Work}
\label{sec:related}

\niparagraph{LLM inference serving.}
Multiple LLM serving systems optimize for their inference performance either through reducing the memory footprint~\cite{arxiv21-lightseq, kvcaching}, improving the kernel execution strategy~\cite{alpa}, determining the partitioning techniques for intra- and inter- operator~\cite{pip, megatronlm, arxiv22-efficiently} execution, or a combination of these~\cite{deepspeed, orca, vllm, hugging-face, tensorflow-serving, tensorrt-llm, nvidia_triton, miao2023spotserve, li2023alpaserve}.
In this work, we specifically tackle the utilization of the current hardware platforms which deploy these models through a compute and I/O suitable platforms, NPUs and PIMs, to create a more efficient system. 
Moreover, to ensure such a heterogeneous system can perform well for LLM inferencing, we offer a scheduling policy.
Prior works that support kernel optimizations for better utilization of GPUs for LLMs, cannot fully mitigate the I/O and bandwidth bottleneck of GEMV kernels. 
Instead in this work, we build a system that can benefit from existing optimizations such as selective batching, KV caching, etc. to offer better utilization across the transformer model architecture.

\niparagraph{PIM for language model support.}
TransPIM~\cite{hpca22-transpim} is a PIM solution that accelerates the end-to-end transformer inference using PIM. 
The work proposes a data loading overhead reduction technique by customizing its dataflow for transformer models.
TransPIM is optimized for attention operations of transformer encoder blocks, making it unsuitable for LLM inference based on decoder blocks.
Furthermore, it is tailored for single-request inference, offering suboptimal performance for batched inference scenarios.
AttAcc~\cite{AttAcc} further offers an accelerator (with PIM) for attention layer to reduce the data movement for the KV matrices.
Instead \neupims proposes a new system for PIM accelerator in addition to the scheduling of operations for the end to end inference of LLMs. 

There are also variety of prior works that leverage PIM for GEMV operations~\cite{hcs23-aquabolt-xl, isscc22-aim-gddr6, micro21-trim, hbm_pim, newton, tensor-dimm, SpaceA} due to their inherent potential in benefits towards bandwidth bound applications.
However, none of these works enable simultaneous execution of PIM and NPU operations, necessary for the efficient execution of LLM inference. 

\niparagraph{Heterogeneous acceleration pipeline for deep learning.}
There are a variety of prior works that propose a pipelined solution for machine learning~\cite{fae, dnnweaver:micro, cosmic:micro, dana}, however none of these prior works leverage PIM to alleviate the bandwidth requirement of LLMs.
Certain prior works use an accelerator for specific models~\cite{hotline, scratchpipe, tabla:hpca}, however, they do not alleviate the under-utilization of GEMV and GEMM operations in transformer decoder blocks.

%

%

\section{Conclusion}
\label{sec:conclusion}

Large Language Model (LLM) inferencing, given its significance, demands dedicated resources that can be deployed at scale. However, these models present a confluence of challenges, encompassing high memory capacity, high compute intensity, and bandwidth constraints.
In this work we propose a novel system, \neupims, that integrates NPU (a general ML accelerator) with PIM technology to mitigate the limitations associated with different operations and their dataflow in the transformer layers.
We introduce a novel scheduling and execution strategy for the proposed system, that can better utilize HBM memory, compute intensive NPU, and the PIM accelerator for LLM inference serving. 
Results indicate that the system developed in this work offers a 1.6$\times$ throughput improvement compared to the baseline system that na\"ively integrates an NPU with the PIM accelerator.
\section*{Acknowledgments}
We thank our shepherd Vidushi Goyal and the anonymous reviewers for their comments and feedback.
This research is supported by Institute of Information \& communications
Technology Planning \& Evaluation (IITP) (No.2022-0-01037, No.2018-0-00503), Information Technology Research Center (ITRC) support program (IITP-2024-2020-0-01795), and Artificial Intelligence Graduate School Program (KAIST) (No.2019-0-00075), funded by the Korea government (MSIT).

\balance

\bibliographystyle{plain}
\bibliography{paper}

\end{document}